\documentclass{article}
\usepackage{bbm}
\usepackage[title,titletoc]{appendix}
\usepackage{graphicx}
\usepackage{amsmath,amssymb,amsthm,amsfonts}
\usepackage{color}
\usepackage{amsfonts}
\usepackage{empheq}
\usepackage{indentfirst}
\usepackage{cite}
\usepackage{mathrsfs}
\usepackage{lineno}%\linenumbers
\usepackage[colorlinks=true, linkcolor=blue, urlcolor=blue]{hyperref}
\allowdisplaybreaks[4]
\newtheorem{thm}{Theorem}[section]
\newtheorem{lem}{Lemma}[section]

\theoremstyle{definition}

\newtheorem{rem}{Remark}[section]

\numberwithin{equation}{section}

\DeclareMathSymbol{\C}{\mathalpha}{AMSb}{"43} \topmargin-.1in

\def\r{\mathbb{R}}
\def\f{\frac}
\def\F{\displaystyle\frac}

\newcommand{\beq}{\begin{equation}}
\newcommand{\eeq}{\end{equation}}
\newcommand{\bea}{\begin{array}}
\newcommand{\eea}{\end{array}}

\newcommand{\eps}{\varepsilon}

\newcommand{\lam}{\lambda}

\newcommand{\alp}{\alpha}

\newcommand{\gam}{\gamma}
\newcommand{\Gam}{\Gamma}
\newcommand{\del}{\delta}

\newcommand{\al}{\alpha}
\newcommand{\be}{\beta}
\def\th{\theta}
\def\Th{\Theta}
\newcommand{\pa}{\partial}

\def\dt{\mathrm{d}t}

\def\d{\mathrm{d}}
\def\p{\mathbb{P}}
\def\ni{\noindent}
\def\a{\mathcal{A}}
\def\b{\mathcal{B}}
\def\c{\mathcal{C}}
\def\proof{{\ni\bf Proof:\quad}}
\def\proofend{{\hfill$\Box$}\\}
\def\pta{{\ni\bf Proof of Theorem \ref{critical curve}: }}
\def\ptb{{\ni\bf Proof of Theorem \ref{properties of critical curve}: }}
\def\ptc{{\ni\bf Proof of Theorem \ref{UV q infty}: }}
\def\ptd{{\ni\bf Proof of Theorem \ref{UV d1 infty}: }}
\def\pte{{\ni\bf Proof of Theorem \ref{UV d2 0}: }}
\def\ptf{{\ni\bf Proof of Theorem \ref{UV d2 infty}: }}

\title{Effects of diffusion and advection on predator prey dynamics in an advective patchy environment}
\author{Qi Wang
\thanks{College of Science, University of Shanghai for Science and Technology,
Shanghai 200093, P.R. China.
Email: \texttt{qwang@usst.edu.cn}.
}}
\date{}

\begin{document}

\maketitle
\begin{abstract}
In this paper,
we consider a specialist predator-prey patchy model over the closed stream network.
We study the dynamics and the asymptotic profiles of positive steady states according to the mortality rate of the specialist predators,
advection and diffusion rates.
We verify that the specialist predators can successfully invade as long as the mortality rate is sufficiently small.
On the other hand,
the impacts of diffusion and advection on the asymptotic profiles of positive steady states and on the concentration of the species are given.
\end{abstract}
\vskip 0.2truein



\section{Introduction}

It is well-known that whether the individuals survive or not in a predominantly unidirectional flow environments (such as streams, rivers, lakes or oceans)
depends on both the environment and the growth of interacting species.
Sometimes, despite the flow induced washout,
the species can always persist in their habitats for many generations.
This phenomenon is so-called "drift paradox" \cite{LNP,LPL,M}.
To explain this paradox,
Speirs and Gurney\cite{SG} proposed a reaction-diffusion-advection equation and indicated that the persistence of single species is possible
when the flow speed is slow relative to the diffusion and the stream is long enough.
Intuitively,
the predominantly unidirectional flow will carry individuals to the downstream end or drive individuals out of the system,
which may be crowded or hostile.
However, random dispersal may drive them to some favorable locations in the upstream \cite{JLL}.
Therefore, one can see
the joint impacts of both undirectional and directed dispersal
rates on the population dynamics of the species are usually complicated and have attracted the attention
of many researchers\cite{HJL,JL,LL,LLM,LML,LPL,SG}.

In real ecological advective environments,
there are various inter-specific relationships including competition, predation.
The two-species reaction-diffusion-advection competition models have been widely studied in \cite{GT,LL,LLL,LNW,LXZ,LZ,MT,TZ,VL1,VL2,YNZ,ZX,ZZ,Z}.
Among them, some results illustrate the conditions under
which the species $u$ or $v$ is stable/unstable.
Various results on the global dynamics
of the two-species reaction-diffusion-advection competition models are also presented.

Regarding the dynamics of two species competition models in a river network,
the authors in \cite{CLW1,CLW2,CSSW2,JLL,JLL2} investigated a two-species Lotka-Volterra competition patchy model over a inland stream.
Let $n\ge 3$, $\boldsymbol{u}=(u_1,u_2,\cdots,u_n)$ and $\boldsymbol{v}=(v_1,v_2,\cdots,v_n)$ be
the population densities of two competing species, respectively, where $u_i$ and $v_i$ are the densities in patch $i$.
Suppose that the dispersal patterns of the individuals and the configuration of the patches are shown
in Fig.\ref{fig:01}

% \begin{figure}[htbp]
% \centering
% {\scalebox{1}[1]{\includegraphics{01.eps}}}
% \caption{\label{fig:01}
% \em A stream with $n$ patches, where $d$ is the random movement rate and $q$ is the directed drift rate. Patch $1$ is the
% upstream end, and patch $n$ is the downstream end.}
% \end{figure}
The competition patchy model over the stream network in Fig. \ref{fig:01} in \cite{CLW1,CLW2,CSSW2,JLL,JLL2} is:
\beq
\label{competiton patch}
\left\{\arraycolsep=1.5pt
\begin{array}{lll}
\F{\d u_i}{\dt}=\sum\limits_{j=1}^{n}(d_1D_{ij}+q_1Q_{ij})u_j+r_iu_i(1-\F{u_i+v_i}{k_i}),\ \ &i\in\{1,2,\cdots,n\},\ t>0,\\[2mm]
\F{\d v_i}{\dt}=\sum\limits_{j=1}^{n}(d_2D_{ij}+q_2Q_{ij})v_j+r_iv_i(1-\F{u_i+v_i}{k_i}),, \   \ &i\in\{1,2,\cdots,n\},\ t>0,\\[2mm]
\boldsymbol{u}(0)=\boldsymbol{u}_0\ge,\not\equiv0,\ \boldsymbol{v}(0)=\boldsymbol{v}_0\ge,\not\equiv0,&
\end{array}
\right.
\eeq
where $d_1,d_2$ are random movement rates;
$q_1,q_2$ represent directed movement rates;
the matrices $D=(D_{ij})$ and $Q=(Q_{ij})$ are the random movement pattern and directed drift pattern of individuals, respectively, where
\beq
\label{DQ}
D=
\begin{pmatrix}
-1&1&0&\cdots&\cdots&0\\
1&-2&1&\cdots&\cdots&0\\
0&\ddots&\ddots&\ddots&\\
\vdots&&\ddots&\ddots&\ddots&\\
0&\cdots&\cdots&1&-2&1\\
0&\cdots&\cdots&\cdots&1&-1
\end{pmatrix},
Q=
\begin{pmatrix}
-1&0&0&\cdots&\cdots&0\\
1&-1&0&\cdots&\cdots&0\\
0&\ddots&\ddots&\ddots&\\
\vdots&&\ddots&\ddots&\ddots&\\
0&\cdots&\cdots&1&-1&0\\
0&\cdots&\cdots&\cdots&1&0
\end{pmatrix}.
\eeq
Recently, Chen, Liu and Wu \cite{CLW2} considered \eqref{competiton patch} with three patches.
They studied if one species is treated as a resident species,
whether the other species can invade or not under
some assumptions on the carrying capacity.

Predation is another common phenomenon in advective environments.
For instance, herbivorous zooplankton and phytoplankton in
water columns.
Compared with the competitive systems,
up to now,
the studies of the evolution of dispersal in predator-prey systems under advective environments are relatively few.
How do the prey persist in advective environments and avoid predations successfully?
How do the predators invade the prey and coexist in advective environments?
To better understand these issues,
Hilker and Lewis \cite{HL} proposed the following predator-prey system in advective environments.
\beq
\left\{\arraycolsep=1.5pt
\begin{array}{lll}
N_t=d_1N_{xx}-qN_{x}+f(N,P)N,\ \ &x\in(0,L),t>0,\\[2mm]
P_t=d_2P_{xx}-\tau qP_{x}+g(N,P)P,\ \ &x\in(0,L),t>0.
\end{array}
\right.
\eeq
Here $N(x,t)$ and $P(x,t)$ are the population densities of the prey and predator species
at time $t$ and location $x$, respectively.
$d_1,d_2$ are dispersal rates,
$q$ denotes the effective advection rate of the prey,
$L$ is the domain length.
The function $f(N,P)$ is given by
\beq
f(N,P)=r_1(x)(1-\F{N}{K_1(x)})-aP,
\eeq
and the function $g(N,P)$ is given by
\beq
\label{g(N,P)}
g(N,P)=\be aN-\gam\ \text{or}\ g(N,P)=r_2(x)(1-\F{P}{K_2(x)})+\be aN,
\eeq
where $r_i(x)(i=1,2)$ and $K_i(x)(i=1,2)$ account for the intrinsic growth rate and the carrying capacity of the prey and predators at location $x$, respectively,
$a$ is the predation rate,
$\be$ is the trophic conversion efficiency, and $\gam$ is the mortality rate of the predators.

By numerical simulations and traveling wave speed approximations,
Hilker and Lewis achieved the coexistence, extinction of both species and survival of one species.
Based on \cite{HL},
some mathematicians studied the dynamics of the predator-prey model in open advective environments further \cite{LN,NLW,NWW,XLN}.
Furthermore,
Wang et al. in \cite{NXS,WN} investigated the dynamics of some predator-prey models in closed advective environments and obtained some significant outcomes.
%by assuming that the effective advection rate of each species is proportional to its diffusion rate.

Motivated by \cite{CLW2,NXS},
in this paper we let $n\ge 3$,
$\boldsymbol{N}=(N_1,N_2,\cdots,N_n),\boldsymbol{P}=(P_1,P_2,\cdots,P_n)$
and consider the following specialist predator-prey patchy model over the closed stream network in Fig.\ref{fig:01}:
\beq
\label{predator prey patch}
\left\{\arraycolsep=1.5pt
\begin{array}{lll}
\F{\d N_i}{\dt}=\sum\limits_{j=1}^{n}(d_1D_{ij}+qQ_{ij})N_j+r_iN_i(1-\F{N_i}{k_i})-aN_iP_i,\ \ &i\in\{1,2,\cdots,n\},\ t>0,\\[2mm]
\F{\d P_i}{\dt}=\sum\limits_{j=1}^{n}(d_2D_{ij}+\tau q Q_{ij})P_j+P_i(\be a N_i-\gam),, \   \ &i\in\{1,2,\cdots,n\},\ t>0,\\[2mm]
\boldsymbol{N}(0)\ge,\neq\boldsymbol{0},\ \boldsymbol{P}(0)\ge,\neq\boldsymbol{0},&
\end{array}
\right.
\eeq
where $N_i$ and $P_i$ are the densities of the prey and predator in patch
$i$ respectively.
Here $r_i(i\in\{1,2,\cdots,n\})$ and $k_i(i\in\{1,2,\cdots,n\})$ account for the intrinsic growth rate and the carrying capacity of the prey %and predators
in patch
$i$, respectively.
The meanings of $a,\be,\gam$ are the same as them in \eqref{g(N,P)}.

The goal of this paper is to study the dynamics of system \eqref{predator prey patch},
and investigate the similarities or differences between \eqref{predator prey patch} and the specialist predator-prey model in \cite{NXS}.

For convenience,
we set
$\a=\{1,2,\cdots,n\}$,
$\b=\{2,\cdots,n-1\}$,
$\c=\{1,\cdots,n-1\}$,
$\boldsymbol{r}=(r_1,r_2,\cdots,r_n)$,
$\boldsymbol{r}+b=(r_1+b,r_2+b,\cdots,r_n+b)$,
$r_{max}=\max\limits_{i\in\a}r_i, r_{min}=\min\limits_{i\in\a}r_i$.
We write $\boldsymbol{r}\ge\boldsymbol{0}(\boldsymbol{r}\gg\boldsymbol{0})$
if $r_i\ge0(r_i>0)$ for
all $i\in\a$,
and $\boldsymbol{r}>\boldsymbol{0}$ if $\boldsymbol{r}\ge\boldsymbol{0}$ and $\boldsymbol{r}\neq\boldsymbol{0}:=(0,\cdots,0)$.

At the end of this section,
we make some assumptions throughout this paper:
\beq
\label{assumption}
\tag{$\mathbf{H}$}
\boldsymbol{r},\boldsymbol{k}\gg\boldsymbol{0},
d_1,d_2,a,\be,\tau>0,q\ge0\ \text{and there exist}\ i\neq j\ \text{such that}\ k_i\neq k_j.
\eeq

The rest of this paper is organized as follows.
In Section \ref{main results},
we state the main results.
In Section \ref{Preliminaries}, we provide some preliminary results.  In Section \ref{proof of the main results},
we show our main results.

\section{Main Results}
\label{main results}

Before describing precisely the outcome of our paper,
we first recall some existing results on the dynamics of the following single-species model:
\beq
\label{single species model}
\left\{\arraycolsep=1.5pt
\begin{array}{ll}
\F{\d N_i}{\dt}=\sum\limits_{j=1}^{n}(d_1D_{ij}+qQ_{ij})N_j+r_iN_i(1-\F{N_i}{k_i}),\ \ &i\in\a,\ t>0,\\[2mm]
\boldsymbol{N}(0)>\boldsymbol{0}.&
\end{array}
\right.
\eeq
It is known in \cite[Lemma 2.2]{CLW2}(see also \cite{LT}) that
system \eqref{single species model} admits a unique positive steady state, denoted by $\boldsymbol{\th}=(\th_1,\cdots,\th_n)\gg0$,
which is globally asymptotically stable for any $d_1>0$ and $q\ge0$.

We are now in a position to state our  main results.

\begin{thm}
\label{critical curve}
Suppose \eqref{assumption} hold.
Let $(\boldsymbol{N}(t),\boldsymbol{P}(t))$ be the solution of \eqref{predator prey patch}.
There exists a critical curve $\gam=\gam^*(q,d_1,d_2)\in(0,+\infty)$ continuously dependent on $q,d_1,d_2$ such that:

$(1)$ if $\gam>\gam^*$,
then $(\boldsymbol{N}(t),\boldsymbol{P}(t))$ converges uniformly
to $(\boldsymbol{\th},\boldsymbol{0})$ as $t\rightarrow+\infty$;

$(2)$ if $\gam<\gam^*$,
then system \eqref{predator prey patch} is uniformly persistent in the sense that
there exists $\eta>0$ such that the solution $(\boldsymbol{N}(t),\boldsymbol{P}(t))$ satisfies
$\liminf\limits_{t\rightarrow+\infty}(N(t))_{min}\ge\eta$, $\liminf\limits_{t\rightarrow+\infty}(P(t))_{min}\ge\eta$.
Moreover, system \eqref{predator prey patch} admits a positive steady state.
\end{thm}

\begin{rem}
Theorem \ref{critical curve} indicates that
the dynamical behavior of system \eqref{predator prey patch} is depicted completely by the critical curve $\gam=\gam^*(q,d_1,d_2)$.
More precisely,
$\gam=\gam^*(q,d_1,d_2)$
classifies the dynamical behavior of system \eqref{predator prey patch} into two scenarios: (i) coexistence; and (ii) persistence of prey only.
Biologically, a large death rate of the predators makes it difficult to invade, while the predators can invade successfully only when the mortality rate of the predators
is less than the critical mortality rate $\gam^*$.
\end{rem}

\begin{thm}
\label{properties of critical curve}
Suppose \eqref{assumption} hold.
The threshold value $\gam^*(q,d_1,d_2)$ satisfies the following properties:

(1) $\lim\limits_{q\rightarrow+\infty}\gam^*(q,d_1,d_2)=\be a k_n$ for $d_1,d_2>0$;

(2) $\lim\limits_{d_1\rightarrow+\infty}\gam^*(q,d_1,d_2)=\F{\be a\sum_{i=1}^n r_i}{\sum_{i=1}^n\f{r_i}{k_i}}$ for $d_2>0$, $q\ge0$;

(3) if $q\ge\max\{r_{max}+2\sqrt{d_1r_{max}},2r_{max}\}$,
then $\lim\limits_{d_1\rightarrow0^+}\gam^*(q,d_1,d_2)=\mu(k_n)$ for $d_2>0$, $q>0$,
where $\mu(k_n)>0$ is the principal eigenvalue of the following system
\beq
\label{mu(k_n)}
\left\{\arraycolsep=1.5pt
\begin {array}{l}
-(d_2+\tau q)\vartheta_1+d_2\vartheta_2=\mu(k_n)\vartheta_1,\\[2mm]
(d_2+\tau q)\vartheta_{j-1}+(-2d_2-\tau q)\vartheta_{j}+d_2\vartheta_{j+1}=\mu(k_n)\vartheta_j,\ j\in\b,\\[2mm]
(d_2+\tau q)\vartheta_{n-1}-d_2\vartheta_n+\be a k_n\vartheta_n=\mu(k_n)\vartheta_n,
\end{array}\right.
\eeq
and $\mu(k_n)$ is strictly increasing with respect to $k_n$;

(4) $\lim\limits_{d_1\rightarrow0^+}\gam^*(0,d_1,d_2)=\lam_1(d_2,0,\be a\boldsymbol{k})>0$ for $d_2>0$,
where $\lam_1(d_2,0,\be a\boldsymbol{k})$ is defined in subsection \ref{The linear eigenvalue problem};

(5) $\lim\limits_{d_2\rightarrow0^+}\gam^*(q,d_1,d_2)
=\max\{\be a\th_1-\tau q,\cdots,\be a\th_{n-1}-\tau q,\be a\th_n\}$,
and
$\lim\limits_{d_2\rightarrow+\infty}\gam^*(q,d_1,d_2)
=\F{\sum_{i=1}^n\be a\th_i}{n}$ for $d_1>0,q\ge0$.
\end{thm}

\begin{rem}
It is easy to see that $\mu(0)=0$.
\end{rem}

Theorem  \ref{properties of critical curve} (1), (2) and (5) indicate that
in closed advective patchy environments $\gam^*(q,d_1,d_2)$ has a positive lower bound for any $q\ge0$ and $d_1,d_2>0$.
This implies that if the prey’s diffusion rate $d_1>0$ is fixed, then the specialist predators can invade successfully as long as their death rate is suitably small.
For a small diffusion rate $d_1$ of the prey, (3) and (4) show that
the specialist predators can always invade successfully in both non-advective and advective environments.
Moreover,
given sufficiently small death rate of the specialist predators,
when the carrying capacity of the prey at patch $n$ increases,
it is more convenient for the predators to invade.
This phenomenon is different from the one in \cite{NXS},
where they verified that
it is difficult for the specialist predators to invade in advective environments,
when $d_1$ is small enough.

Next, we investigate the asymptotic profiles of the positive steady state $(\boldsymbol{N}^*,\boldsymbol{P}^*)$.

\begin{thm}
\label{UV q infty}
Suppose \eqref{assumption} hold,
and $0<\gam<\be ak_n$.
Let $(\boldsymbol{N}^*,\boldsymbol{P}^*)$ be a positive steady state of \eqref{predator prey patch}.
Then it satisfies
$(N^*_i,P^*_i)\rightarrow(0,0)$ uniformly for $i\in\c$ as $q\rightarrow+\infty$,
and $(N^*_n,P^*_n)\rightarrow(\F{\gam}{\be a},\F{1}{a}(r_n-\F{r_n}{k_n}\F{\gam}{\be a}))$ as $q\rightarrow+\infty$.
\end{thm}

\begin{thm}
\label{UV d1 infty}
Suppose \eqref{assumption} hold, $0<\gam<\be ak_{min}$.
Let $(\boldsymbol{N}^*,\boldsymbol{P}^*)$ be a positive steady state of \eqref{predator prey patch}.
Then it satisfies
\beq
\label{12}
(N^*_i,P^*_i)\rightarrow(\F{\gam}{\be a},(1+\F{\tau q}{d_2})^{i-1}\F{\sum_{j=1}^nr_j(1-\f{\gam}{\be ak_j})}{\f{d_2a}{\tau q}[(1+\f{\tau q}{d_2})^n-1]})
\eeq
uniformly for $i\in\a$ as $d_1\rightarrow+\infty$.
\end{thm}

\begin{thm}
\label{UV d2 0}
Suppose \eqref{assumption} hold,
$0<\gam<\gam^*(q,d_1,d_2)$,
and $q>\max\big\{r_{max}+2\sqrt{d_1 r_{max}},\F{2\be aN^*_n}{\tau}\}$.
Let $(\boldsymbol{N}^*,\boldsymbol{P}^*)$ be a positive steady state of \eqref{predator prey patch}.
Then it satisfies
$(N^*_i,P^*_n)\rightarrow(\th_i,0)$ as $d_2\rightarrow0^+$.
Moreover,
\beq
\max\limits_{i\in\c}\Big(P^*_i-P^*_n(\F{d_1}{d_2+\tau q})^{n-i}\Big)\rightarrow0\ \text{as}\ d_2\rightarrow0^+.
\eeq
\end{thm}

\begin{thm}
\label{UV d2 infty}
Suppose \eqref{assumption} hold, $0<\gam<\F{\be a\sum_i^n\th_i}{n}$.
Let $(\boldsymbol{N}^*,\boldsymbol{P}^*)$ be a positive steady state of \eqref{predator prey patch}.
Then it satisfies
$(N^*_i,P^*_n)\rightarrow(\hat{\th}_i,z)$ as $d_2\rightarrow+\infty$,
where $(\hat{\th}_i,z)$ satisfies
\beq
\left\{\arraycolsep=1.5pt
\begin{array}{ll}
\sum\limits_{j=1}^{n}(d_1D_{ij}+qQ_{ij})\hat{\th}_j+\hat{\th}_i(r_i-\F{r_i}{k_i}\hat{\th}_i-az)=0,\ \ &i\in\a,\\[2mm]
\be a\sum\limits_{i=1}^{n}\hat{\th}_i=\gam n, \   \ &i\in\a.
\end{array}
\right.
\eeq
\end{thm}

%Note that the condition $\max\limits_{i\in\a}b_ik_i=b_nk_n$
%in Theorem \ref{UV q infty}, \ref{UV d2 0} is technique.
%Whether it can be deleted or not is unknown up to now.
%We hope to consider this issue in our future studies.

Theorem \ref{UV q infty} implies that
sufficiently large advection will affect the distribution of species in the network.
More precisely,
two species will coexist and concentrate at the patch $n$
if the advection rate is large enough.
Theorems \ref{UV d1 infty} and \ref{UV d2 infty} show that
two species can coexist for large diffusion rates.
%Meanwhile, large diffusion rates result in the spatial homogeneous distribution of species.
Theorem \ref{properties of critical curve}(5) and \ref{UV d2 0} illustrate the
successful invasion of the specialist predators
once diffusion rate $d_2$ is enough small,
but their population density tends to zero as the diffusion rate tends to zero.
This phenomenon is partially due to the lack of other food sources of the specialist predators except the prey.
Another reason is that the specialist predators cannot catch up with the prey when their diffusion rate is sufficiently small.

\section{Preliminaries}
\label{Preliminaries}

In this section, we aim to exhibit some fundamental results, which will be utilized in later
sections.

\subsection{The linear eigenvalue problem}
\label{The linear eigenvalue problem}

We first consider an auxiliary linear eigenvalue problem.
\beq
\label{linear eigenvalue problem}
\sum\limits_{j=1}^{n}(d D_{ij}+qQ_{ij})\phi_j+h_i\phi_i=\lam\phi_i,\ \ i\in\a,
\eeq
where $d>0,q\ge0$.
It is well-known from \cite{CLW1,CLW2} that
problem \eqref{linear eigenvalue problem} admits a unique principal eigenvalue $\lam_1(d,q,\boldsymbol{h})=s(dD+qQ+diag(h_i))$(which is the spectral bound of $dD+qQ+diag(h_i)$) associated with a nonnegative eigenvector $\boldsymbol{\phi}=(\phi_1,\cdots,\phi_n)>0$,
since the matrix $dD+qQ+diag(h_i)$ is irreducible and essentially nonnegative for any $d>0,q\ge0$,
where $D$ and $Q$ are defined by \eqref{DQ}.

\begin{lem}
\label{prop of lam1}
Let $D$ and $Q$ be defined in \eqref{DQ},
$d,q>0$,
and $\lam_1(d,q,\boldsymbol{h})$ be the principal
eigenvalue of \eqref{linear eigenvalue problem}.
Suppose that $\boldsymbol{h}\gg\boldsymbol{0}$ and there exist $i\neq j$ such that $h_i\neq h_j$.
Then, we have the following:

(i) for each fixed $d>0$,
$\lam_1(d,q,\boldsymbol{h})$ is strictly decreasing with respect to $q$ in $[0,+\infty)$ when $h_1\ge\cdots\ge h_n$ with at least one strict inequality,
and $\lam_1(d,q,\boldsymbol{h})$ is strictly increasing with respect to $q$ when $h_1\le\cdots\le h_n$ with at least one strict inequality;

(ii) $\lim\limits_{q\rightarrow+\infty}\lam_1(d,q,\boldsymbol{h})=h_n$;

(iii) $\lim\limits_{d\rightarrow0}\lam_1(d,q,\boldsymbol{h})=\max\{h_1-q,\cdots,h_{n-1}-q,h_n\}$\ \text{and}\ $\lim\limits_{d\rightarrow+\infty}\lam_1(d,q,\boldsymbol{h})=\F{\sum_{i=1}^{n}h_i}{n}$;

(iv) if $\boldsymbol{h}>\boldsymbol{g}$,
then $\lam_1(d,q,\boldsymbol{h})\ge\lam_1(d,q,\boldsymbol{g})$,
and the equality holds only if $\boldsymbol{h}=\boldsymbol{g}$.
\end{lem}

\proof
The proofs of (i)-(iii) are standard and we refer to \cite{CLW1} for the proofs.
We now show the statement (iv).
Indeed, we can rewrite \eqref{linear eigenvalue problem} as the following forms:
\beq
\label{linear eigenvalue problem1}
\left\{\arraycolsep=1.5pt
\begin {array}{l}
-(d+q)\phi_1+d\phi_2+h_1\phi_1=\lam_1\phi_1,\\[2mm]
(d+q)\phi_{j-1}+(-2d-q)\phi_{j}+d\phi_{j+1}+h_j\phi_j=\lam_1\phi_j,\ j\in\b,\\[2mm]
(d+q)\phi_{n-1}-d\phi_n+h_n\phi_n=\lam_1\phi_n.
\end{array}\right.
\eeq
Differentiating \eqref{linear eigenvalue problem1} with respect to $h_1$,
we have
\beq
\label{01}
\left\{\arraycolsep=1.5pt
\begin {array}{l}
-(d+q)\F{\pa\phi_1}{\pa h_1}+d\F{\pa\phi_2}{\pa h_1}+h_1\F{\pa\phi_1}{\pa h_1}+\phi_1=\F{\pa\lam_1}{\pa h_1}\phi_1+\lam_1\F{\pa\phi_1}{\pa h_1},\\[2mm]
(d+q)\F{\pa\phi_{j-1}}{\pa h_1}+(-2d-q)\F{\pa\phi_j}{\pa h_1}+d\F{\pa\phi_{j+1}}{\pa h_1}+h_j\F{\pa\phi_j}{\pa h_1}=\F{\pa\lam_1}{\pa h_1}\phi_j+\lam_1\F{\pa\phi_j}{\pa h_1},\ j\in\b,\\[2mm]
(d+q)\F{\pa\phi_{n-1}}{\pa h_1}-d\F{\pa\phi_n}{\pa h_1}+h_n\F{\pa\phi_n}{\pa h_1}=\F{\pa\lam_1}{\pa h_1}\phi_n+\lam_1\F{\pa\phi_n}{\pa h_1}.
\end{array}
\right.
\eeq
Multiplying the first(second,third) equation of \eqref{01} by $\phi_1$($\phi_j,\phi_n$) and the first(second,third) equation of \eqref{linear eigenvalue problem1} by $\F{\pa\phi_1}{\pa h_1}$($\F{\pa\phi_j}{\pa h_1},\F{\pa\phi_n}{\pa h_1}$),
and subtracting each other,
it then follows that
\beq
\label{02}
\left\{\arraycolsep=1.5pt
\begin {array}{l}
d(\F{\pa\phi_2}{\pa h_1}\phi_1-\phi_2\F{\pa\phi_1}{\pa h_1})+\phi_1^2=\F{\pa\lam_1}{\pa h_1}\phi_1^2,\\[2mm]
(d+q)(\F{\pa\phi_{j-1}}{\pa h_1}\phi_j-\phi_{j-1}\F{\pa\phi_j}{\pa h_1})+d(\F{\pa\phi_{j+1}}{\pa h_1}\phi_j-\phi_{j+1}\F{\pa\phi_j}{\pa h_1})=\F{\pa\lam_1}{\pa h_1}\phi_j^2,\ j\in\b,\\[2mm]
(d+q)(\F{\pa\phi_{n-1}}{\pa h_1}\phi_n-\phi_{n-1}\F{\pa\phi_n}{\pa h_1})=\F{\pa\lam_1}{\pa h_1}\phi_n^2.
\end{array}\right.
\eeq
Let $b_i=(\F{d}{d+q})^{i-1}$.
Then we have that $b_id=b_{i+1}(d+q)$.
This implies that
\beq
\label{03}
\bea{l}
b_1d(\F{\pa\phi_2}{\pa h_1}\phi_1-\phi_2\F{\pa\phi_1}{\pa h_1}\phi_1')
+b_2(d+q)(\F{\pa\phi_1}{\pa h_1}\phi_2-\phi_{1}\F{\pa\phi_2}{\pa h_1})\\
+b_2d(\F{\pa\phi_3}{\pa h_1}\phi_2-\phi_{3}\F{\pa\phi_2}{\pa h_1})+b_3(d+q)(\F{\pa\phi_2}{\pa h_1}\phi_3-\phi_{2}\F{\pa\phi_3}{\pa h_1})\\
+b_3d(\F{\pa\phi_4}{\pa h_1}\phi_3-\phi_{4}\F{\pa\phi_3}{\pa h_1})+b_4(d+q)(\F{\pa\phi_3}{\pa h_1}\phi_4-\phi_{3}\F{\pa\phi_4}{\pa h_1})\\
+\cdots\\
+b_{n-1}d(\F{\pa\phi_n}{\pa h_1}\phi_{n-1}-\phi_{n}\F{\pa\phi_{n-1}}{\pa h_1})+b_n(d+q)(\F{\pa\phi_{n-1}}{\pa h_1}\phi_n-\phi_{n-1}\F{\pa\phi_n}{\pa h_1})\\
=0.
\eea
\eeq
Then by \eqref{02} and \eqref{03},
there holds
\beq
\F{\pa\lam_1}{\pa h_1}(b_1\phi_1^2+\cdots+b_n\phi_n^2)=b_1\phi_1^2>0.
\eeq
Therefore, $\F{\pa\lam_1}{\pa h_1}>0$.
Similarly, one can have that $\F{\pa\lam_1}{\pa h_i}>0$ for all $i\in\a$.
The proof is completed.
\proofend

\begin{lem}
\label{hi=h0}
Let $D$ and $Q$ be defined in \eqref{DQ},
$d>0,q\ge0$,
and $\lam_1(d,q,\boldsymbol{h})$ be the principal
eigenvalue of \eqref{linear eigenvalue problem}.
Suppose that $\boldsymbol{h}\gg\boldsymbol{0}$.
If $h_1=h_2=\cdots=h_n:=h_0$,
then $\lam_1(d,q,\boldsymbol{h})=h_0$ for all $d>0$, $q\ge0$.
\end{lem}
Indeed this lemma can be deduced directly,
we omit the proof here.

\subsection{Properties of the positive steady state of single species model}

Recall that for $d_1>0$ and $q>0$,
\eqref{single species model} possesses a unique positive steady state $\boldsymbol{\th}=(\th_1,\cdots,\th_n)\gg0$,
which satisfies
\beq
\label{single ss}
\sum\limits_{l=1}^{n}(d_1D_{il}+qQ_{il})\th_l+r_i\th_i(1-\F{\th_i}{k_i})=0,\ \ i\in\a,
\eeq
i.e.
\beq
\label{theta}
\left\{\arraycolsep=1.5pt
\begin {array}{l}
-(d_1+q)\th_1+d_1\th_2+r_1\th_1(1-\F{\th_1}{k_1})=0,\\[2mm]
(d_1+q)\th_{j-1}+(-2d_1-q)\th_{j}+d_1\th_{j+1}+r_j\th_j(1-\F{\th_j}{k_j})=0,\ j\in\b,\\[2mm]
(d_1+q)\th_{n-1}-d_1\th_n+r_n\th_n(1-\F{\th_n}{k_n})=0.
\end{array}\right.
\eeq
Then we collect some useful results for $\boldsymbol{\th}$.

\begin{lem}
\label{prop of th}
Suppose $d_1>0,q\ge0$.
Then we have

(i) $\F{\min_{i\in\a}b_ik_i}{b_i}\le\th_i\le\F{\max_{i\in\a}b_ik_i}{b_i}$, where $b_i=(\F{d}{d+q})^{i-1}$;

(ii) if $k_1\le k_2\le\cdots\le k_n$,
then $\th_1\le \th_2\le\cdots\le \th_n$;

(iii) if $q\ge r_{max}+2\sqrt{d_1r_{max}}$,
then $\th_i$ is strictly increasing with respect to $i$;

(iv) if $q\ge\max\{r_{max}+2\sqrt{d_1r_{max}},2r_{max}\}$,
then for $i\in\c$
\beq
\label{inequality of thi}
\th_n(\F{d_1+q+\f{2r_{max}}{k_{min}}\th_n}{d_1})^{i-n}<\th_i<\th_n(\F{d_1+q-2r_{max}}{d_1})^{i-n}.
\eeq
Moreover,
\beq
\label{lim of thn}
\lim\limits_{q\rightarrow+\infty}\th_n=\lim\limits_{d_1\rightarrow0}\th_n=k_n,
\eeq
and
\beq
\label{lim of thi}
\lim\limits_{q\rightarrow+\infty}\max\limits_{i\in\a}(\th_i-\th_n(\F{d_1}{d_1+q})^{n-i})=\lim\limits_{d_1\rightarrow0}\max\limits_{i\in\a}(\th_i-\th_n(\F{d_1}{d_1+q})^{n-i})=0;
\eeq

(v) $\lim\limits_{d_1\rightarrow+\infty}\th_i=\F{\sum_{i=1}^n r_i}{\sum_{i=1}^n\f{r_i}{k_i}}$;

(vi) for $d_1>0$ small, if $q=0$,
then $\max\limits_{i\in\a}|\th_i-k_i|<C_*d_1$
where $C_*=\F{2\max_{i\in\a}(\sum_{l=1}^nD_{il}k_l)}{r_{min}}$.
\end{lem}

\proof
(i) Let $b_i=(\F{d_1}{d_1+q})^{i-1}$ and $w_i=b_i\th_i$($i\in\a$).
We see that $w_i$ satisfies
\beq
\label{w}
\left\{\arraycolsep=1.5pt
\begin {array}{l}
(d_1+q)(w_2-w_1)+r_1(1-\F{w_1}{b_1k_1})w_1=0,\\[2mm]
d_1w_{j-1}+(-2d_1-q)w_j+(d_1+q)w_{j+1}+r_j(1-\F{w_j}{b_jk_j})w_j=0,\ j\in\b,\\[2mm]
d_1(w_{n-1}-w_{n})+r_n(1-\F{w_n}{b_nk_n})w_n=0.
\end{array}\right.
\eeq
If $w_1=\max\limits_{i\in\a}w_i$,
then $w_2-w_1\le0$,
which implies from the first equation of \eqref{w} that
$w_i\le w_1\le b_1k_1$.
Similarly, if $w_n=\max\limits_{i\in\a}w_i$,
then $w_i\le w_n\le b_n k_n$.
If $w_{i_0}=\max\limits_{i}w_i$ for some $i_0\in\b$,
then $w_{i_0-1}-2w_{i_0}+w_{i_0+1}\le0$.
The second equation of \eqref{w} then yields that $w_i\le w_{i_0}\le b_{i_0}k_{i_0}$.
Therefore, $w_i\le\max\limits_{i\in\a}b_ik_i$
and $\th_i\le\F{\max_{i\in\a}b_ik_i}{b_i}$.
Another inequality $\F{\min_{i\in\a}b_ik_i}{b_i}\le\th_i$ can be obtained similarly.

(ii) Set $a_l=\F{\th_{l+1}-\th_l}{\th_l}$($l\in\c$).
Then following equations can be deduced
directly by \eqref{theta}.
\beq
\label{a}
\left\{\arraycolsep=1.5pt
\begin {array}{l}
d_1a_1-q+r_1(1-\F{\th_1}{k_1})=0,\\[2mm]
d_1\big[(a_j-a_{j-1})\th_j+a_{j-1}^2\th_{j-1}\big]-qa_{j-1}\th_{j-1}+r_j\th_j(1-\F{\th_j}{k_j})=0,\ j\in\b,\\[2mm]
q-d_1a_{n-1}+r_n\F{\th_n}{\th_{n-1}}(1-\F{\th_n}{k_n})=0.
\end{array}\right.
\eeq

Suppose that $a_1\le0$.
By the first equation in \eqref{a},
we conclude that $k_1>\th_1$.
Then $k_2\ge k_1>\th_1\ge\th_2$.
This implies that
$1-\F{\th_2}{k_2}>0$
and further $(a_2-a_1)\th_2+a_1^2\th_1<0$ by the second equation of \eqref{a} with $j=2$.
Thus $a_2<a_1\le0$,
which implies that
$\th_3<\th_2<k_2\le k_3$,
$1-\F{\th_3}{k_3}>0$.
By the second equation of \eqref{a} with $j=3$,
we arrive at $a_3<a_2$.
Thus, by induction,
it follows that
\beq
\label{15}
a_{n-1}<\cdots<a_1\le0,
\eeq
which indicates that
\beq
\th_n<\th_{n-1}<\cdots<\th_2\le\th_1<k_1\le\cdots\le k_n.
\eeq
However, the third equation of \eqref{a} yields that
\beq
d_1a_{n-1}=q+r_n\F{\th_n}{\th_{n-1}}(1-\F{\th_n}{k_n})>0,
\eeq
a contradiction to \eqref{15}.
Therefore, $a_1>0$.

Suppose that $a_2\le 0$.
Using the second equation in \eqref{theta} with $j=2$(which is equivalent to $r_2\th_2(1-\F{\th_2}{k_2})+d_1a_2\th_2-(d_1+q)a_1\th_1=0$),
we arrive at $k_3\ge k_2>\th_2\ge\th_3$.
It then follows from the second equation in \eqref{a} with $j=3$ that
$a_3<a_2\le0$.
Thus $\th_4<\th_3\le\th_2<k_2\le k_3\le k_4$.
By induction,
there holds
\beq
a_{n-1}<\cdots<a_2\le0,
\eeq
and furthermore
$\th_n<\cdots<\th_3\le\th_2<k_n$,
which contradicts to the third equation in \eqref{a}.
Hence $a_2>0$.
By the similar methods as above,
we reach that $\th_n>\cdots>\th_1$.

(iii)
By a transformation $v_i=(\F{d_1}{d_1+q})^{\f{i-1}{2}}\th_i$,
we can get that
\beq
\label{v}
\left\{\arraycolsep=1.5pt
\begin {array}{l}
\sqrt{d_1(d_1+q)}(v_2-v_1)+(r_1(1-\F{\th_1}{k_1})-d_1-q+\sqrt{d_1(d_1+q)})v_1=0,\\[2mm]
\sqrt{d_1(d_1+q)}(v_{j-1}-2v_j+v_{j+1})+(r_j(1-\F{\th_j}{k_j})-2d_1-q+2\sqrt{d_1(d_1+q)})v_j=0,\ j\in\b,\\[2mm]
(d_1+q)\th_{n-1}-d_1\th_n+r_n\th_n(1-\F{\th_n}{k_n})=0.
\end{array}\right.
\eeq
Since $q\ge r_{max}+2\sqrt{d_1r_{max}}$,
we see from the first and second equations of \eqref{v} that $v_1<v_2<\cdots<v_n$.
Therefore,
$\th_1<\th_2<\cdots<\th_n$.

(iv) Set $\Th_i=\th_n(\F{d_1+q-2r_{max}}{d_1})^{i-n}$.
By direct calculations,
we obtain from $q\ge 2r_{max}$ that
\beq
\bea{l}
\sum\limits_{l=1}^n(d_1D_{1l}+qQ_{1l})\Th_l+r_1\Th_1(1-\F{\th_1}{k_1})=-(d_1+q)\Th_1+d_1\Th_2+r_1\Th_1(1-\F{\th_1}{k_1})\\[2mm]
=\th_n[(-d_1-q)(\F{d_1+q-2r_{max}}{d_1})^{1-n}+d_1(\F{d_1+q-2r_{max}}{d_1})^{2-n}+r_1(\F{d_1+q-2r_{max}}{d_1})^{1-n}(1-\F{\th_1}{k_1})]\\[2mm]
<\th_n(\F{d_1+q-2r_{max}}{d_1})^{1-n}(-d_1-q+d_1(\F{d_1+q-2r_{max}}{d_1})+r_{max})\le0,
\eea
\eeq
and
\beq
\begin {array}{l}
\sum\limits_{l=1}^n(d_1D_{jl}+qQ_{jl})\Th_l+r_j\Th_j(1-\F{\th_j}{k_j})=(d_1+q)\Th_{j-1}+(-2d_1-q)\Th_{j}+d_1\Th_{j+1}+r_j\Th_j(1-\F{\th_j}{k_j})\\[2mm]
=\th_n(\F{d_1+q-2r_{max}}{d_1})^{j-n-1}[\F{q-2r_{max}}{d_1}(d_1(\F{d_1+q-2r_{max}}{d_1})-d_1-q)\\[2mm]
+r_1(\F{d_1+q-2r_{max}}{d_1})-r_1(\F{d_1+q-2r_{max}}{d_1})\F{\th_j}{k_j}],\\[2mm]
<\th_n(\F{d_1+q-2r_{max}}{d_1})^{j-n-1}[d_1+q+(r_{max}-2d_1-q)(\F{d_1+q-2r_{max}}{d_1})\\[2mm]
+d_1(\F{d_1+q-2r_{max}}{d_1})^2]\le0,\ j\in\b.
\end{array}
\eeq
Now, set
$u_i=\Th_i-\th_i$ satisfying
\beq
\sum\limits_{l=1}^n(d_1D_{jl}+qQ_{jl})u_l+r_ju_j(1-\F{\th_j}{k_j})<0,\ j\in\c.
\eeq
Clearly, $u_n=0$.
By a further transformation $u_i=z_i(\F{d_1}{d_1+q})^{\f{1-i}{2}}$,
we get that
\beq
\label{04}
\left\{\arraycolsep=1.5pt
\begin {array}{l}
\sqrt{d_1(d_1+q)}(z_2-z_1)+(r_1(1-\F{\th_1}{k_1})-d_1-q+\sqrt{d_1(d_1+q)})z_1<0,\\[2mm]
\sqrt{d_1(d_1+q)}(z_{j-1}-2z_j+z_{j+1})+(r_j(1-\F{\th_j}{k_j})-2d_1-q+2\sqrt{d_1(d_1+q)})z_j<0,\ j\in\b.
\end{array}\right.
\eeq
If there exists $i_0\in\b$ such that $0\ge z_{i_0}=\min\limits_{i\in\a}z_i$,
then $(z_{i_0-1}-2z_{i_0}+z_{i_0+1})\ge0$.
Since $q\ge r_{max}+2\sqrt{d_1r_{max}}$ leads to $r_{i_0}(1-\F{\th_{i_0}}{k_{i_0}})-2d_1-q+2\sqrt{d_1(d_1+q)}<0$,
then there holds
\beq
\sqrt{d_1(d_1+q)}(z_{i_0-1}-2z_{i_0}+z_{i_0+1})+\big(r_{i_0}(1-\F{\th_{i_0}}{k_{i_0}})-2d_1-q+2\sqrt{d_1(d_1+q)}\big)z_{i_0}\ge0,
\eeq
which is a contradiction.
Hence, $z_i>\min\{0,z_1\}$ for $i\in\c$.
If $z_1<0$, then $z_2>z_1$.
By $q\ge r_{max}+2\sqrt{d_1r_{max}}$,
there holds
\beq
\sqrt{d_1(d_1+q)}(z_2-z_1)+(r_1(1-\F{\th_1}{k_1})-d_1-q+\sqrt{d_1(d_1+q)})z_1>0,
\eeq
which contradicts to \eqref{04}.
Hence $z_1\ge0$,
and therefore $z_i>0$,
that is,
$\th_i<\Th_i$ for $i\in\c$.
This proves the second inequality in \eqref{inequality of thi}.

Next, set $\vartheta_i=\th_n(\F{d_1+q+\f{2r_{max}}{k_{min}}\th_n}{d_1})^{i-n}$.
Then we have from $q\ge r_{max}+2\sqrt{d_1r_{max}}$ that
\beq
\bea{l}
\sum\limits_{l=1}^n(d_1D_{il}+qQ_{il})
\vartheta_i+r_i\vartheta_i(1-\F{\th_i}{k_i})
>\sum\limits_{l=1}^n(d_1D_{il}+qQ_{il})\vartheta_i-r_{max}\vartheta_i\F{\th_n}{k_{min}}>0.
\eea
\eeq
By the same arguments,
the first inequality in \eqref{inequality of thi} also holds.

A direct summation of the equation of $\th_i$ gives
\beq
\label{05}
\sum\limits_{i=1}^{n}r_i\th_i=\sum\limits_{i=1}^{n}r_i\F{\th_i^2}{k_i}.
\eeq
It then follows from \eqref{inequality of thi} that
\beq
\th_n^2\min\limits_{i\in\a}\F{r_i}{k_i}\sum\limits_{i=1}^{n}(\F{d_1+q+\f{2r_{max}}{k_{min}}\th_n}{d_1})^{2i-2n}\le r_{max}\th_n\sum\limits_{i=1}^{n}(\F{d_1+q-2r_{max}}{d_1})^{i-n},
\eeq
which implies that
\beq
\min\limits_{i\in\a}\F{r_i}{k_i}\th_n\F{1-(\f{d_1}{d_1+q+\f{2r_{max}}{k_{min}}\th_n})^{2n}}{1-(\f{d_1}{d_1+q+\f{2r_{max}}{k_{min}}\th_n})^2}<\F{r_{max}}{1-\f{d_1}{d_1+q-2r_{max}}}.
\eeq
Obviously,
for sufficiently large $q$ or sufficiently small $d_1$,
we have that
\beq
1-(\f{d_1}{d_1+q+\f{2r_{max}}{k_{min}}\th_n})^{2n}\ge\F12,\
1-(\f{d_1}{d_1+q+\f{2r_{max}}{k_{min}}\th_n})^{2}>0,\ \text{and}\ 1-\f{d_1}{d_1+q-2r_{max}}\ge\F12.
\eeq
Hence, we can deduce that
\beq
\label{thn is bounded}
\min\limits_{i\in\a}\F{r_i}{k_i}\th_n<4r_{max},\ \text{for}\ q\gg1\ \text{or}\ d_1\ll1.
\eeq
Next, we show the desired limit.
Dividing \eqref{05} by $\th_n$,
we then observe that
\beq
\th_n\sum\limits_{i=1}^{n}\F{r_i}{k_i}(\F{\th_i}{\th_n})^2=\sum\limits_{i=1}^{n}r_i\F{\th_i}{\th_n}.
\eeq
Since
\beq
(\F{d_1+q+\f{2r_{max}}{k_{min}}\th_n}{d_1})^{i-n}\le\F{\th_i}{\th_n}\le(\F{d_1+q-2r_{max}}{d_1})^{i-n},
\eeq
then we see that
\beq
\lim\limits_{q\rightarrow+\infty}\F{\th_i}{\th_n}=\lim\limits_{d_1\rightarrow0}\F{\th_i}{\th_n}=0,\ \lim\limits_{q\rightarrow+\infty}\F{\th_i}{\th_n}=\lim\limits_{d_1\rightarrow0}(\F{\th_i}{\th_n})^2=0,\ \text{for}\ i\in\c,
\eeq
which, together with \eqref{thn is bounded},
allows us to obtain the limit \eqref{lim of thn}.

In view of \eqref{inequality of thi},
we reach that
\beq
\th_n((\F{d_1}{d_1+q+\f{2r_{max}}{k_{min}}\th_n})^{n-i}-(\F{d_1}{d_1+q})^{n-i})\le\th_i-\th_n(\F{d_1}{d_1+q})^{n-i}\le\th_n((\F{d_1}{d_1+q-2r_{max}})^{n-i}-(\F{d_1}{d_1+q})^{n-i}).
\eeq
Therefore,
the limit \eqref{lim of thi} is correct.

(v)
By statement (i),
we have
that $\th_i$ is uniformly bounded for $i\in\a$ as $d_1\rightarrow+\infty$.
Dividing the system \eqref{theta} by $d_1$
leads to
\beq
\label{07}
\left\{\arraycolsep=1.5pt
\begin {array}{l}
\th_2-\th_1-\F{q}{d_1}\th_1+\F{r_1}{d_1}\th_1(1-\F{\th_1}{k_1})=0,\\[2mm]
(1+\F{q}{d_1})\th_{j-1}+(-2-\F{q}{d_1})\th_{j}+\th_{j+1}+\F{r_j}{d_1}\th_j(1-\F{\th_j}{k_j})=0,\ j\in\b,\\[2mm]
(1+\F{q}{d_1})\th_{n-1}-\th_n+\F{r_n}{d_1}\th_n(1-\F{\th_n}{k_n})=0.
\end{array}\right.
\eeq
Then we have that
\beq
\th_1=\th_2=\cdots=\th_n:=\rho\ge0,
\eeq
as $d_1\rightarrow+\infty$.
By applying the statement (i) again,
one can get that $\rho>0$.
Adding \eqref{07} up,
we get
\beq
\sum\limits_{i=1}^n r_i\th_i(1-\F{\th_i}{k_i})=0,
\eeq
which means that $\lim\limits_{d_1\rightarrow+\infty}\th_i=\rho=\F{\sum_{i=1}^nr_i}{\sum_{i=1}^n\f{r_i}{k_i}}$.

(vi) The idea of the proof of this statement is to get a sequence of appropriate upper and lower solutions in terms
of $d_1$.
For each $d_1>0$,
letting $\bar{\th}_i=k_i+C_*d_1$,
we compute
\beq
\bea{l}
\sum\limits_{l=1}^{n}(d_1D_{il}\bar{\th}_l)+\F{r_i}{k_i}\bar{\th}_i(k_i-\bar{\th}_i)=d_1(\sum\limits_{l=1}^{n}D_{il}k_l-C_*r_i(1+C_*\F{d_1}{k_i}))\\
\le d_1(\sum\limits_{l=1}^{n}D_{il}k_l-C_*r_i)\le0,
\eea
\eeq
i.e. $\bar{\boldsymbol{\th}}$ is an upper solution.

Setting $\underline{\th}_i=k_i-C_*d_1$,
we see that for $d_1>0$ small, $\underline{\th}_i>0$.
Then we compute, for $d_1>0$ small
\beq
\bea{l}
\sum\limits_{l=1}^{n}(d_1D_{il}\underline{\th}_l)+\F{r_i}{k_i}\underline{\th}_i(k_i-\underline{\th}_i)=d_1(\sum\limits_{l=1}^{n}D_{il}k_l+C_*r_i-C_*^2\F{d_1r_i}{k_i})\\
=d_1(\sum\limits_{l=1}^{n}D_{il}k_l+\F12 C_*r_i+C_*r_i(\F12-C_*\F{d_1}{k_i}))\ge d_1(\sum\limits_{l=1}^{n}D_{il}k_l+\F12 C_*r_i)\ge0,
\eea
\eeq
i.e. $\underline{\boldsymbol{\th}}$ is a lower solution.
Clearly, $\bar{\th}_i>\underline{\th}_i$.
Therefore, by the similar arguments in the proof of \cite[Lemma 2.2]{CLW2}(see also \cite{LT}),
we have that $\bar{\th}_i>\th_i>\underline{\th}_i$,
and statement (vi) is
correct.
This ends the proof.
\proofend

%\begin{rem}
%\label{thi<kn}
%If $\max\limits_{i\in\a}b_ik_i=b_nk_n$,
%it directly follows from Lemma \ref{prop of th}(i) that $\th_i\le k_n$.
%\end{rem}

\subsection{Maximum Principle}

In this subsection,
we shall introduce some important maximum/comparison principles.

\begin{lem}[Strong maximum principle I]
\label{Strong maximum principle for elliptic}
Let $\boldsymbol{w}$ satisfies
\beq
\label{elliptic wi}
\left\{\arraycolsep=1.5pt
\begin {array}{l}
-(d+q)w_1+(d+q)w_2+c_1w_1\ge0,\\[2mm]
d w_{j-1}+(-2d-q)w_{j}+(d+q)w_{j+1}+c_jw_j\ge0,\ j\in\b,\\[2mm]
d w_{n-1}-d w_n+c_nw_n\ge0,
\end{array}
\right.
\eeq
where $c_i\in\r$.
If $\boldsymbol{w}\le\boldsymbol{0}$,
then either $\boldsymbol{w}\ll\boldsymbol{0}$ or $\boldsymbol{w}=\boldsymbol{0}$.
\end{lem}

\proof
Set $\Sigma=\{i\in\a:w_i=0\}$.
We need to show $\Sigma=\{1,2,\cdots,n\}$
if $\Sigma\neq\emptyset$.
If $w_1=0$,
then we may find from the first equation of \eqref{elliptic wi} and $\boldsymbol{w}\le\boldsymbol{0}$ that
$w_2=0$.
Furthermore, the second and the third equations of \eqref{elliptic wi} imply that $w_3=\cdots=w_n=0$.
Similarly,
we arrive at $\boldsymbol{w}=\boldsymbol{0}$ provided that $w_n=0$.
If now $w_1<0,w_2=0$,
by $w_3\le0=w_2$,
we then have from the second equation of \eqref{elliptic wi} that
\beq
0\le d w_{1}+(-2d-q)w_{2}+(d+q)w_{3}+c_2w_2=d (w_{1}-w_2)+(d+q)(w_{3}-w_2)<0,
\eeq
which is a contradiction.
This implies that $w_1$ can not be negative if $w_2=0$.
That is, $w_2=0$ yields that $w_1=w_2=0$.
By the similar method,
one can get that $\boldsymbol{w}=\boldsymbol{0}$ provided that there exists a $i_0\in\a$ such that $w_{i_0}=0$.
\proofend

\begin{lem}[Strong maximum principle II]
\label{Strong maximum principle for elliptic II}
Let $c_i\in\r$, $\boldsymbol{u}$ satisfy
\beq
\label{ui}
\sum\limits_{j=1}^{n}(d D_{ij}+qQ_{ij})u_j+c_iu_i\le0,\ \ i\in\a,
\eeq
where the inequality above is strict for some $j_0\in\a$.
Suppose that $\lam_1(d,q,\boldsymbol{c})<0$.
Then there holds $\boldsymbol{u}\gg\boldsymbol{0}$.
\end{lem}

\proof
Indeed, after setting $w_i=(\F{d}{d+q})^{i-1}u_i$,
\eqref{ui} is equivalent to
\beq
\left\{\arraycolsep=1.5pt
\begin {array}{l}
-(d+q)w_1+(d+q)w_2+c_1w_1\le0,\\[2mm]
d w_{j-1}+(-2d-q)w_{j}+(d+q)w_{j+1}+c_jw_j\le0,\ j\in\b,\\[2mm]
d w_{n-1}-d w_n+c_nw_n\le0.
\end{array}
\right.
\eeq
We shall verify this lemma in several steps.

\emph{Step 1.
Let $\boldsymbol{\phi}>\boldsymbol{0}$ be the non-negative eigenvector corresponding to $\lam_1(d,q,\boldsymbol{c})$.
Then $\boldsymbol{\phi}\gg0$.
}

Clearly, it follows from \eqref{linear eigenvalue problem} that $\bar{\phi}_i:=(\F{d}{d+q})^{i-1}\phi_i$ satisfies
\beq
\left\{\arraycolsep=1.5pt
\begin {array}{l}
-(d+q)\bar{\phi}_1+(d+q)\bar{\phi}_2+c_1\bar{\phi}_1=\lam_1(d,q,\boldsymbol{c})\bar{\phi}_1\le0,\\[2mm]
d \bar{\phi}_{j-1}+(-2d-q)\bar{\phi}_{j}+(d+q)\bar{\phi}_{j+1}+c_j\bar{\phi}_j=\lam_1(d,q,\boldsymbol{c})\bar{\phi}_j\le0,\ j\in\b,\\[2mm]
d \bar{\phi}_{n-1}-d \bar{\phi}_n+c_n\bar{\phi}_n=\lam_1(d,q,\boldsymbol{c})\bar{\phi}_n\le0,
\end{array}
\right.
\eeq
and the inequality above is strict for some $j_0\in\a$.
Furthermore, Lemma \ref{Strong maximum principle for elliptic} yields that $\boldsymbol{\bar{\phi}}\gg\boldsymbol{0}$,
and hence $\boldsymbol{\phi}\gg\boldsymbol{0}$.

\emph{Step 2. $\boldsymbol{w}\gg\boldsymbol{0}$.}
Suppose that there exists a $i_0\in\a$ such that $w_{i_0}=\min\limits_{i\in\a}w_i<0$.
Then we can find some $\xi>0$ such that $\boldsymbol{w}+\xi\boldsymbol{\bar{\phi}}\ge\boldsymbol{0}$.
In fact, we can define $\xi_0=\inf\{\xi>0:w_i+\xi\bar{\phi}_i\ge0\ \forall i\in\a\}>0$.
Therefore $\boldsymbol{w}+\xi_0\boldsymbol{\bar{\phi}}\ge\boldsymbol{0}$.
Note that $\boldsymbol{w}+\xi_0\boldsymbol{\bar{\phi}}$ satisfies
\beq
\left\{\arraycolsep=1.5pt
\begin {array}{l}
-(d+q)(w_1+\xi_0\bar{\phi}_1)+(d+q)(w_2+\xi_0\bar{\phi}_2)+c_1(w_1+\xi_0\bar{\phi}_1)\le0,\\[2mm]
d (w_{j-1}+\xi_0\bar{\phi}_{j-1})+(-2d-q)(w_j+\xi_0\bar{\phi}_j)+(d+q)(w_{j+1}+\xi_0\bar{\phi}_{j+1})+c_j(w_j+\xi_0\bar{\phi}_j)\le0,\ j\in\b,\\[2mm]
d (w_{n-1}+\xi_0\bar{\phi}_{n-1})-d (w_n+\xi_0\bar{\phi}_n)+c_n(w_n+\xi_0\bar{\phi}_n)\le0,
\end{array}
\right.
\eeq
and the inequality above is also strict for some $j_0\in\a$.
This implies that $\boldsymbol{w}+\xi_0\boldsymbol{\bar{\phi}}=\boldsymbol{0}$ can not happen.
We than conclude from Lemma \ref{Strong maximum principle for elliptic} that $\boldsymbol{w}+\xi_0\boldsymbol{\bar{\phi}}\gg\boldsymbol{0}$,
which contradicts the definition of $\xi_0$.
Therefore, $\boldsymbol{w}\ge\boldsymbol{0}$.
Combined with Lemma \ref{Strong maximum principle for elliptic} again,
we reach that $\boldsymbol{w}\gg\boldsymbol{0}$.
That is, $\boldsymbol{u}\gg\boldsymbol{0}$.
\proofend

\begin{lem}[Comparison principle]
\label{Comparison principle}
Let $\boldsymbol{w}(t)$ satisfy
\beq
\label{let wi}
\left\{\arraycolsep=1.5pt
\begin {array}{l}
\F{\d w_1}{\dt}\ge-(d+q)w_1+(d+q)w_2+c_1w_1,\ 0<t<T,\\[2mm]
\F{\d w_j}{\dt}\ge d w_{j-1}+(-2d-q)w_{j}+(d+q)w_{j+1}+c_iw_i,\ j\in\b,\ 0<t<T,\\[2mm]
\F{\d w_n}{\dt}\ge d w_{n-1}-d w_n+c_nw_n,\ 0<t<T,\\[2mm]
\boldsymbol{w}(0)>\boldsymbol{0},
\end{array}\right.
\eeq
where $c_i\in\r$.
Then $\boldsymbol{w}(t)\ge\boldsymbol{0}$.
\end{lem}

\proof
Consider the case where $c_i\le0$ for every $i\in\a,$
and the strict inequality
\beq
\label{wt>}
\F{\d w_i}{\dt}>\sum\limits_{j=1}^{n}(d D_{ij}+q Q_{ij})w_j+c_i w_i
\eeq
firstly.
If there exists a point $(i_0,t_0)\in\a\times(0,T]$ such that $w_{i_0}(t_0)=\inf\limits_{(i,t)\in\a\times(0,T)}w_i(t)<0$,
\eqref{wt>} then yields that
\beq
0\ge\F{\d w_{i_0}}{\dt}>\sum\limits_{j=1}^{n}(d D_{i_0j}+q Q_{i_0j})w_j+c_{i_0} w_{i_0}\ge0,
\eeq
which is a contradiction.
This contradiction shows that $\boldsymbol{w}(t)\ge\boldsymbol{0}$.

For the case
\beq
\F{\d w_i}{\dt}\ge\sum\limits_{j=1}^{n}(d D_{ij}+q Q_{ij})w_j+c_i w_i,
\eeq
write $w^{\eps}_i(t)=w_i(t)+\eps t$ where $\eps>0$.
Then
$\F{\d w^{\eps}_i}{\dt}>\sum\limits_{j=1}^{n}(d D_{ij}+q Q_{ij})w^{\eps}_j+c_i w^{\eps}_i$
and so $w^{\eps}_i\ge0$.
Let $\eps\rightarrow0$ to find $w_i\ge0$.
This proves the theorem for the case $c_i\le0$.
For arbitrary $c_i$,
we choose $b_i\ge c_i$
and define $v_i(t)=e^{-b_it}w_i(t)$.
It is easy to see from \eqref{let wi} that
$v_i(t)$ satisfies
\beq
\left\{\arraycolsep=1.5pt
\begin {array}{l}
\F{\d v_1}{\dt}\ge-(d+q)v_1+(d+q)v_2+(c_1-b_1)v_1,\ 0<t<T,\\[2mm]
\F{\d v_j}{\dt}\ge d v_{j-1}+(-2d-q)v_{j}+(d+q)v_{j+1}+(c_i-b_i)v_i,\ j\in\b,\ 0<t<T,\\[2mm]
\F{\d v_n}{\dt}\ge d v_{n-1}-d v_n+(c_n-b_n)v_n,\ 0<t<T,\\[2mm]
\boldsymbol{v}(0)>\boldsymbol{0}.
\end{array}\right.
\eeq
Since $c_i-b_i\le0$,
the above conclusion for $c_i\le0$ implies that $\boldsymbol{v}\ge\boldsymbol{0}$.
The conclusion $\boldsymbol{w}(t)\ge\boldsymbol{0}$ follows from $w_i(t)=e^{b_it}v_i(t)$.
\proofend

\begin{lem}[Strong maximum principle III]
\label{Strong maximum principle}
Let $\boldsymbol{w}(t)$ satisfy
\beq
\left\{\arraycolsep=1.5pt
\begin {array}{l}
\F{\d w_1}{\dt}\le-(d+q)w_1+(d+q)w_2+c_1w_1,\ 0<t<T,\\[2mm]
\F{\d w_j}{\dt}\le d w_{j-1}+(-2d-q)w_{j}+(d+q)w_{j+1}+c_iw_i,\ j\in\b,\ 0<t<T,\\[2mm]
\F{\d w_n}{\dt}\le d w_{n-1}-d w_n+c_nw_n,\ 0<t<T,\\[2mm]
\boldsymbol{w}(0)<\boldsymbol{0},
\end{array}\right.
\eeq
where $c_i\in\r$.
If $w_{i_0}(t_0)=\max\limits_{i\in\a,t>0}w_i(t)=M\ge0$ for some $t_0>0,i_0\in\a$,
then $w_i(t)\equiv M$ for all $t<t_0$.
\end{lem}

\proof
Without loss of generality,
we may assume that $c_i\le0$,
otherwise one can set $v_i(t)=e^{-b_it}w_i(t)$ for $b_i\ge c_i$ and consider $v_i(t)$.
Suppose by contradiction that there exists $(i_1,t_1)\in\a\times(0,t_0)$ such that $w_{i_1}(t_1)\le M_1<M$.
Set $u_i(t)=M-w_i(t)-(M-M_1)he^{-\alp(t-t_1)}$,
where $h$ is a small positive constant.
We then have that
\beq
\sum\limits_{j=1}^{n}(dD_{ij}+qQ_{ij})u_j+c_iu_i-\F{\d u_i}{\dt}\le-e^{\alp(t-t_1)}(M-M_1)(\alp h+c_ih+\sum\limits_{j=1}^{n}(dD_{ij}+qQ_{ij})h)<0,
\eeq
provided $\alp\gg1$.
Since $u_i(t_1)=M-w_i(t_1)-(M-M_1)h\ge(1-h)(M-M_i)>0$ for $0<h\ll 1$.
It then follows from Lemma \ref{Comparison principle} that $\boldsymbol{u}(t)\ge\boldsymbol{0}$.
Furthermore, $w_i(t)\le M-(M-M_1)he^{-\alp(t-t_1)}<M$ for $t>t_1$.
This implies that $w_{i_0}(t_0)<M$,
which is a contradiction.
The proof of this lemma is hence finished.
\proofend

\section{Proof of the main results}
\label{proof of the main results}

This section is devoted to the proofs of our main results.

\subsection{Dynamic classification in the $q-\gam$ plane}

Clearly, the steady states of \eqref{predator prey patch} consists of the trivial steady state $(\boldsymbol{0},\boldsymbol{0})$, the semi-trivial steady state $(\boldsymbol{\th},\boldsymbol{0})$ and the positive steady state $(\boldsymbol{N}^*,\boldsymbol{P}^*)$.

\begin{lem}
Suppose that \eqref{assumption} hold.
The trivial steady state $(\boldsymbol{0},\boldsymbol{0})$ of \eqref{predator prey patch} is always unstable.
\end{lem}

\proof
The linearization eigenvalue problem of\eqref{predator prey patch} at $(\boldsymbol{0},\boldsymbol{0})$ is

\beq
\left\{
\bea{l}
\sum\limits_{j=1}^{n}(d_1 D_{ij}+qQ_{ij})\phi_j+r_i\phi_i=\lam\phi_i,\ i\in\a,\\[2mm]
\sum\limits_{j=1}^{n}(d_2 D_{ij}+\tau qQ_{ij})\psi_j-\gam\psi_i=\lam\psi_i\ i\in\a.
\eea
\right.
\eeq
Note by $\boldsymbol{r}\gg\boldsymbol{0}$ that $\lam_1(d_1,q,\boldsymbol{r})>0$.
We  conclude that $(\boldsymbol{0},\boldsymbol{0})$ is always unstable.
\proofend

\begin{lem}
Suppose that \eqref{assumption} hold.
There exists a unique continuous critical curve $\Gam:\gam=\gam^*(q,d_1,d_2)$ such that $(\boldsymbol{\th},\boldsymbol{0})$ is locally asymptotically stable if $\gam>\gam^*(q,d_1,d_2)$ and unstable if $0<\gam<\gam^*(q,d_1,d_2)$.
\end{lem}

\proof
The linearization eigenvalue problem of \eqref{predator prey patch} at $(\boldsymbol{\th},\boldsymbol{0})$ is
\beq
\left\{
\bea{l}
\sum\limits_{j=1}^{n}(d_1 D_{ij}+qQ_{ij})\phi_j+r_i(1-\F{2\th_i}{k_i})\phi_i-a\th_i\psi_i=\mu\phi_i,\ i\in\a,\\[2mm]
\sum\limits_{j=1}^{n}(d_2 D_{ij}+\tau qQ_{ij})\psi_j+(\be a\th_i-\gam)\psi_i=\mu\psi_i\ i\in\a.
\eea
\right.
\eeq
Clearly, the local stability of $(\boldsymbol{\th},\boldsymbol{0})$ is determined by the sign of the principal eigenvalue of $L\psi_i=\sum\limits_{j=1}^{n}(d_2 D_{ij}+\tau qQ_{ij})\psi_j+(\be a\th_i-\gam)\psi_i$; namely, $(\boldsymbol{\th},\boldsymbol{0})$ is locally asymptotically stable if $\lam_1(d_2,\tau q,\be a\boldsymbol{\th}-\gam)<0$ and unstable if $\lam_1(d_2,\tau q,\be a\boldsymbol{\th}-\gam)>0$.
Furthermore, we have
\beq
\lam_1(d_2,\tau q,\be a\boldsymbol{\th}-\gam)=\lam_1(d_2,\tau q,\be a\boldsymbol{\th})-\gam.
\eeq
Therefore,
we let $\gam^*(q,d_1,d_2)=\lam_1(d_2,\tau q,\be a\boldsymbol{\th}).$ Then it follows from Lemma \ref{prop of lam1}(iv) that
\beq
\label{gam*}
\left\{
\bea{l}
\lam_1(d_2,\tau q,\be a\boldsymbol{\th}-\gam)>0\ \text{if}\ 0<\gam<\gam^*\\
\lam_1(d_2,\tau q,\be a\boldsymbol{\th}-\gam)=0\ \text{if}\ \gam=\gam^*\\
\lam_1(d_2,\tau q,\be a\boldsymbol{\th}-\gam)<0\ \text{if}\ \gam>\gam^*.
\eea
\right.
\eeq
\proofend

\pta
We shall split the proof into several steps.

\emph{Step 1.}
Let $U_i(t)=(\F{d_1}{d_1+q})^{i-1}N_i(t)$ and $V_i(t)=(\F{d_2}{d_2+\tau q})^{i-1}P_i(t)$,
where $(\boldsymbol{N}(t),\boldsymbol{P}(t))$ is the solution of \eqref{predator prey patch}.
Then $(\boldsymbol{U}(t),\boldsymbol{V}(t))$ satisfies
\beq
\label{parabolic system UV}
\left\{\arraycolsep=1.5pt
\begin {array}{l}
\F{\d U_1}{\dt}=-(d_1+q)U_1+(d_1+q)U_2+r_1U_1(1-\F{N_1}{k_1})-aU_1P_1,\\[2mm]
\F{\d U_j}{\dt}=d_1 U_{j-1}+(-2d_1-q)U_{j}+(d_1+q)U_{j+1}+r_jU_j(1-\F{N_j}{k_j})-aU_jP_j,\ j\in\b,\\[2mm]
\F{\d U_n}{\dt}=d_1U_{n-1}-d_1U_n+r_nU_n(1-\F{N_n}{k_n})-aU_nP_n,\\[2mm]
\F{\d V_1}{\dt}=-(d_2+\tau q)V_1+(d_2+\tau q)V_2+V_1(\be a N_1-\gam),\\[2mm]
\F{\d V_j}{\dt}=d_2 V_{j-1}+(-2d_2-\tau q)V_{j}+(d_2+\tau q)V_{j+1}+V_j(\be a N_j-\gam),\ j\in\b,\\[2mm]
\F{\d V_n}{\dt}=d_2V_{n-1}-d_2V_n+V_n(\be a N_n-\gam),\\[2mm]
\boldsymbol{U}(0)=(1,\F{d_1}{d_1+q},\cdots,(\F{d_1}{d_1+q})^{n-1})\cdot\boldsymbol{N}(0)>\boldsymbol{0},\\[2mm]
\boldsymbol{V}(0)=(1,\F{d_2}{d_2+\tau q},\cdots,(\F{d_2}{d_2+\tau q})^{n-1})\cdot\boldsymbol{P}(0)>\boldsymbol{0}.
\end{array}\right.
\eeq
Clearly, Lemma \ref{Strong maximum principle} leads to $\boldsymbol{U},\boldsymbol{V}\gg\boldsymbol{0}$ since $\boldsymbol{U}(0),\boldsymbol{V}(0)>\boldsymbol{0}$,
and then $\boldsymbol{N},\boldsymbol{P}\gg\boldsymbol{0}$.
Note that
\beq
\label{Ui<=}
\left\{\arraycolsep=1.5pt
\begin {array}{l}
\F{\d U_1}{\dt}\le-(d_1+q)U_1+(d_1+q)U_2+r_1U_1(1-\F{U_1}{k_1}),\\[2mm]
\F{\d U_j}{\dt}\le d_1 U_{j-1}+(-2d_1-q)U_{j}+(d_1+q)U_{j+1}+r_jU_j(1-\F{U_j}{k_j}(\F{d_1+q}{d_1})^{j-1}),\ j\in\b,\\[2mm]
\F{\d U_n}{\dt}\le d_1U_{n-1}-d_1U_n+r_nU_n(1-\F{U_n}{k_n}(\F{d_1+q}{d_1})^{n-1}).
\end{array}\right.
\eeq
Let $C_1=\max\limits_{i\in\a}\{k_i,U_i(0)\}$, and $\mathcal{U}_i=U_i-C_1$.
It then follows that
$\mathcal{U}_i$ satisfies $\mathcal{U}_i(0)\le0$ and
\beq
\left\{\arraycolsep=1.5pt
\begin {array}{l}
\F{\d\mathcal{U}_1}{\dt}\le-(d_1+q)\mathcal{U}_1+(d_1+q)\mathcal{U}_2+r_1C_1(-\F{\mathcal{U}_1}{k_1}),\\[2mm]
\F{\d\mathcal{U}_j}{\dt}\le d_1\mathcal{U}_{j-1}+(-2d_1-q)\mathcal{U}_{j}+(d_1+q)\mathcal{U}_{j+1}+r_jC_1(-\F{\mathcal{U}_j}{k_j}(\F{d_1+q}{d_1})^{j-1}),\ j\in\b,\\[2mm]
\F{\d\mathcal{U}_n}{\dt}\le d_1\mathcal{U}_{n-1}-d_1\mathcal{U}_n+r_nC_1(-\F{\mathcal{U}_n}{k_n}(\F{d_1+q}{d_1})^{n-1}).
\end{array}\right.
\eeq
From the above inequalities,
one can get that $\mathcal{U}_i\le0$ for $i\in\a$ and $t\ge0$,
and hence $U_i\le C_1$, $0<N_i\le C_1(\F{d_1+q}{d_1})^{n-1}$ for $t>0$.

\emph{Step 2.}
Next, we verify $V_i$ and $P_i$ are bounded uniformly for $i\in\a$, $t>0$.
We rewrite system \eqref{parabolic system UV} as
\beq
\label{rewrite para system UV}
\left\{\arraycolsep=1.5pt
\begin {array}{l}
\F{\d U_1}{\dt}=-(d_1+q)U_1+(d_1+q)U_2+r_1U_1(1-\F{N_1}{k_1})-aU_1P_1,\\[2mm]
(\F{d_1+q}{d_1})^{j-1}\F{\d U_j}{\dt}=(d_1 U_{j-1}+(-2d_1-q)U_{j}+(d_1+q)U_{j+1}+r_jU_j(1-\F{N_j}{k_j})-aU_jP_j)(\F{d_1+q}{d_1})^{j-1},\ j\in\b,\\[2mm]
(\F{d_1+q}{d_1})^{n-1}\F{\d U_n}{\dt}=(d_1U_{n-1}-d_1U_n+r_nU_n(1-\F{U_n}{k_n})-aU_nP_n)(\F{d_1+q}{d_1})^{n-1},\\[2mm]
\F{\d V_1}{\dt}=-(d_2+\tau q)V_1+(d_2+\tau q)V_2+V_1(\be a N_1-\gam),\\[2mm]
(\F{d_2+\tau q}{d_2})^{j-1}\F{\d V_j}{\dt}=(d_2 V_{j-1}+(-2d_2-\tau q)V_{j}+(d_2+\tau q)V_{j+1}+V_j(\be a N_j-\gam))(\F{d_2+\tau q}{d_2})^{j-1},\ j\in\b,\\[2mm]
(\F{d_2+\tau q}{d_2})^{n-1}\F{\d V_n}{\dt}=(d_2V_{n-1}-d_2V_n+V_n(\be a N_n-\gam))(\F{d_2+\tau q}{d_2})^{n-1},\\[2mm]
\boldsymbol{U}(0)>\boldsymbol{0},\\[2mm]
\boldsymbol{V}(0)>\boldsymbol{0}.
\end{array}\right.
\eeq
Let $w(t)=\sum\limits_{i=1}^{n}(\be(\F{d_1+q}{d_1})^{i-1}U_i+(\F{d_2+\tau q}{d_2})^{i-1}V_i)$.
Adding all the equations in \eqref{rewrite para system UV} up
yields that
\beq
w'(t)+\gam w(x)\le\be(r_{max}+\gam)\sum\limits_{i=1}^{n}(\F{d_1+q}{d_1})^{i-1}U_i\le\be(r_{max}+\gam)C_1n(\F{d_1+q}{d_1})^{n-1}.
\eeq
By Gronwall inequality,
we have
\beq
w(t)\le w(0)e^{-\gam t}+\F{\be(r_{max}+\gam)C_1 n(\F{d_1+q}{d_1})^{n-1}}{\gam}(1-e^{-\gam t}),
\eeq
which implies that $w(t)$ is bounded for all $t>0$.
Then one can get that $0<V_i\le(\F{d_2+\tau q}{d_2})^{i-1}V_i< w(t)$ is also bounded uniformly.
Hence $P_i(t)$ is bounded uniformly and the statement (i) holds.

\emph{Step 3. Consider the case $\gam>\gam^*$.}

\emph{Step 3.1.}
Let $\boldsymbol{f}(t)$ be the solution of
\beq
\left\{\arraycolsep=1.5pt
\begin {array}{l}
\F{\d f_1}{\dt}=-(d_1+q)f_1+(d_1+q)f_2+r_1f_1(1-\F{f_1}{k_1}),\\[2mm]
\F{\d f_j}{\dt}=d_1 f_{j-1}+(-2d_1-q)f_{j}+(d_1+q)f_{j+1}+r_jf_j(1-\F{f_j}{k_j}(\F{d_1+q}{d_1})^{j-1}),\ j\in\b,\\[2mm]
\F{\d f_n}{\dt}=d_1f_{n-1}-d_1f_n+r_nf_n(1-\F{f_n}{k_n}(\F{d_1+q}{d_1})^{n-1}),\\[2mm]
\boldsymbol{f}(0)=(1,\F{d_1}{d_1+q},\cdots,(\F{d_1}{d_1+q})^{n-1})\cdot\boldsymbol{N}(0)>\boldsymbol{0}.
\end{array}\right.
\eeq
The comparison principle (Lemma \ref{Comparison principle}) and \eqref{Ui<=} yield that $U_i\le f_i$.
Observe from \cite[Lemma 2.2]{CLW2} that $\lim\limits_{t\rightarrow+\infty}f_i=(\F{d_1}{d_1+q})^{i-1}\th_i$.
This implies that
\beq
\label{limsup Ui}
\limsup\limits_{t\rightarrow+\infty}U_i\le(\F{d_1}{d_1+q})^{i-1}\th_i\ \text{for all}\ i\in\a.
\eeq
Then for any $\eps>0$,
there exists $T_1>0$ such that $N_i(t)<\th_i+\eps$ for all $t\ge T_1$.
Let $\boldsymbol{g}(t)$ be the solution of
\beq
\label{mathscr Vi}
\left\{\arraycolsep=1.5pt
\begin {array}{l}
\F{\d g_1}{\dt}=-(d_2+\tau q)g_1+(d_2+\tau q)g_2+g_1(\be a(\th_1+\eps)-\gam),\\[2mm]
\F{\d g_j}{\dt}=d_2 g_{j-1}+(-2d_2-\tau q)g_{j}+(d_2+\tau q)g_{j+1}+g_j(\be a(\th_j+\eps)-\gam),\ j\in\b,\\[2mm]
\F{\d g_n}{\dt}=d_2g_{n-1}-d_2g_n+g_n(\be a(\th_n+\eps)-\gam),\\[2mm]
\boldsymbol{g}(T_1)=\boldsymbol{V}(T_1).
\end{array}\right.
\eeq
Clearly,
one can get that $V_i(t)\le g_i(t)$ for all $t\ge T_1$ by using Lemma \ref{Comparison principle}.
Since $\lam_1(d_2,\tau q,\be a\boldsymbol{\th}-\gam)<0$ when $\gam>\gam^*$,
there exists $\eps>0$ small enough such that $\lam_1(d_2,\tau q,\be a(\boldsymbol{\th}+\boldsymbol{\eps})-\gam)<0$.
Since it is easy to see that $C(\F{d_2}{d_2+\tau q})^{i-1}e^{\lam_1(d_2,\tau q,\be a(\boldsymbol{\th}+\boldsymbol{\eps})-\gam)(t-T_1)}\boldsymbol{\phi}$
is a super solution of \eqref{mathscr Vi} for $C$ large enough,
where $\boldsymbol{\phi}$ is the principal eigenfunction of the eigenvalue problem \eqref{linear eigenvalue problem1} with respect to $\lam_1(d_2,\tau q,\be a(\boldsymbol{\th}+\boldsymbol{\eps})-\gam)$.
Lemma \ref{Comparison principle} then yields that
\beq
0\le\limsup\limits_{t\rightarrow+\infty}V_i(t)\le\lim\limits_{t\rightarrow+\infty}C\phi_i e^{\lam_1(d_2,\tau q,\be a(\boldsymbol{\th}+\boldsymbol{\eps})-\gam)(t-T_1)}=0.
\eeq

\emph{Step 3.2.}
Since $\lim\limits_{t\rightarrow+\infty}V_i(t)=0$,
therefore, $\lim\limits_{t\rightarrow+\infty}P_i(t)=0$ and
for any $\eps>0$,
there exists $T_2>T_1$ such that $P_i(t)\le\eps$ for $t\ge T_2$,
which leads to
\beq
\left\{\arraycolsep=1.5pt
\begin {array}{l}
\F{\d U_1}{\dt}\ge-(d_1+q)U_1+(d_1+q)U_2+r_1U_1(1-\F{N_1}{k_1})-aU_1\eps,\\[2mm]
\F{\d U_j}{\dt}\ge d_1 U_{j-1}+(-2d_1-q)U_{j}+(d_1+q)U_{j+1}+r_jU_j(1-\F{N_j}{k_j})-aU_j\eps,\ j\in\b,\\[2mm]
\F{\d U_n}{\dt}\ge d_1U_{n-1}-d_1U_n+r_nU_n(1-\F{N_n}{k_n})-aU_n\eps.
\end{array}\right.
\eeq
Let $\boldsymbol{f^{\eps}}(t)$ be the solution of
\beq
\label{mathscr Ueps}
\left\{\arraycolsep=1.5pt
\begin {array}{l}
\F{\d f^{\eps}_1}{\dt}=-(d_1+q)f^{\eps}_1+(d_1+q)f^{\eps}_2+r_1f^{\eps}_1(1-\F{f^{\eps}_1}{k_1})-af^{\eps}_1\eps,\\[2mm]
\F{\d f^{\eps}_j}{\dt}=d_1 f^{\eps}_{j-1}+(-2d_1-q)f^{\eps}_{j}+(d_1+q)f^{\eps}_{j+1}+r_jf^{\eps}_j(1-\F{N_j}{k_j}\F{d_1+q}{d_1})^{j-1})-af^{\eps}_j\eps,\ j\in\b,\\[2mm]
\F{\d f^{\eps}_n}{\dt}=d_1f^{\eps}_{n-1}-d_1f^{\eps}_n+r_nf^{\eps}_n(1-\F{N_n}{k_n}\F{d_1+q}{d_1})^{n-1})-af^{\eps}_n\eps,\\[2mm]
\boldsymbol{f}(T_2)=\boldsymbol{U}(T_2).
\end{array}\right.
\eeq
Lemma \ref{Comparison principle} implies that $U_i(t)\ge f^{\eps}_i(t)$ for all $t\ge T_2$.
By applying \cite[Lemma 2.2]{CLW2} and
choosing $\eps>0$ sufficiently small,
we have that $\lim\limits_{t\rightarrow+\infty}f^{\eps}_i(t)=\th^{\eps}_i(\F{d_1}{d_1+q})^{i-1}$ uniformly for $i\in\a$,
where $\th^{\eps}_i(\F{d_1}{d_1+q})^{i-1}$ is the unique positive steady state
solution of \eqref{mathscr Ueps}.
Furthermore, by
virtue of Lemma \ref{prop of th}(i) and \eqref{mathscr Ueps},
we can obtain that $0\le \th^{\eps}_i\le(\F{d_1+q}{d_1})^{n-1}\max\limits_{i\in\a}k_i(\F{d_1}{d_1+q})^{i-1}$.
Note that we can deduce that $\lim\limits_{\eps\rightarrow0}\th^{\eps}_i=\th_i$. That is
\beq
\liminf\limits_{t\rightarrow+\infty}U_i(t)\ge\th_i(\F{d_1}{d_1+q})^{i-1}.
\eeq
Combining with \eqref{limsup Ui},
we get that
\beq
\lim\limits_{t\rightarrow+\infty}U_i(t)=\th_i(\F{d_1}{d_1+q})^{i-1},\ \lim\limits_{t\rightarrow+\infty}N_i(t)=\th_i.
\eeq
Hence the statement (ii) holds.

\emph{Step 4. Consider the case $\gam<\gam^*$.}

\emph{Step 4.1.}
System \eqref{predator prey patch} is uniformly persistent.

In order to prove the uniform persistence of system \eqref{predator prey patch}(or the equivalent system \eqref{parabolic system UV}) in the case where $\gam<\gam^*$,
let $\Theta(t)$ be the solution
semiflow generated by system \eqref{parabolic system UV} on the state space $\p$,
where
\beq
\p=\{(\boldsymbol{U},\boldsymbol{V})\in\r^n\times\r^n: \boldsymbol{U}>\boldsymbol{0}, \boldsymbol{V}>\boldsymbol{0}\}.
\eeq
Define
\beq
\p_0=\{(\boldsymbol{U},\boldsymbol{V})\in\p: \boldsymbol{U}\neq\boldsymbol{0}\ \text{and}\ \boldsymbol{V}\neq\boldsymbol{0}\},
\eeq
and $\pa \p_0=\p\setminus\p_0$.
Let
\beq
M_{\pa}=\{(\boldsymbol{U}(0),\boldsymbol{V}(0))\in\pa\p_0:\Theta(t)(\boldsymbol{U}(0),\boldsymbol{V}(0))\in\pa\p_0, \forall t\ge 0\},
\eeq
and $\omega((\boldsymbol{U}(0),\boldsymbol{V}(0)))$ be the omega limit set of the forward orbit $\delta^+((\boldsymbol{U}(0),\boldsymbol{V}(0)))=\{\Theta(t)(\boldsymbol{U}(0),\boldsymbol{V}(0)):t\ge 0\}$.
By Lemma \ref{Strong maximum principle},
we can conclude that $\p_0$ is open in $\p$ and forward invariant under the dynamics generated by system \eqref{parabolic system UV},
and $\pa\p_0$ contains steady state points $(\boldsymbol{0},\boldsymbol{0})$, $(\boldsymbol{\th},\boldsymbol{0})$.
We claim that
\beq
\bigcup_{(\boldsymbol{U}(0),\boldsymbol{V}(0))\in M_{\pa}}\omega((\boldsymbol{U}(0),\boldsymbol{V}(0)))\in\{(\boldsymbol{0},\boldsymbol{0}),(\boldsymbol{\th},\boldsymbol{0})\}.
\eeq
In fact,
for any $(\boldsymbol{U}(0),\boldsymbol{V}(0))\in M_{\pa}$,
we have $\Theta(t)(\boldsymbol{U}(0),\boldsymbol{V}(0))\in\pa\p_0, \forall t\ge 0$.
This implies that $\boldsymbol{U}(t;(\boldsymbol{U}(0),\boldsymbol{V}(0)))=\boldsymbol{0}$
or $\boldsymbol{V}(t;(\boldsymbol{U}(0),\boldsymbol{V}(0)))=\boldsymbol{0}$ for $t\ge 0$.
If $\boldsymbol{U}(t;(\boldsymbol{U}(0),\boldsymbol{V}(0)))=\boldsymbol{0}$ for $t\ge 0$,
then $\boldsymbol{P}(t;(\boldsymbol{N}(0),\boldsymbol{P}(0)))$ satisfies
\beq
\left\{\arraycolsep=1.5pt
\begin {array}{l}
\F{\d P_1}{\dt}=-(d_2+\tau q)P_1+d_2P_2-\gam P_1,\\[2mm]
\F{\d P_j}{\dt}=(d_2+\tau q)P_{j-1}+(-2d_2-\tau q)P_{j}+d_2P_{j+1}-\gam P_j,\ j\in\b,\\[2mm]
\F{\d P_n}{\dt}=(d_2+\tau q)P_{n-1}-d_2P_n-\gam P_n,\\[2mm]
\boldsymbol{P}(0)>\boldsymbol{0}.
\end{array}\right.
\eeq
It follows that $\F{\d (e^{\gam t}\sum_{i=1}^n P_i)}{\dt}=0$,
which hence implies that
$0\le\lim\limits_{t\rightarrow+\infty}P_i=0$.
In the case where
$\boldsymbol{U}(\tau_0;(\boldsymbol{U}(0),\boldsymbol{V}(0)))\neq\boldsymbol{0}$ for some $\tau_0>0$,
we have $\boldsymbol{U}(t;(\boldsymbol{U}(0),\boldsymbol{V}(0)))>\boldsymbol{0}$ for all $t>\tau_0$ by Lemma \ref{Strong maximum principle},
which implies that
$\boldsymbol{V}(t;(\boldsymbol{U}(0),\boldsymbol{V}(0)))=\boldsymbol{0}$ for all $t>\tau_0$.
Thus $\boldsymbol{N}(t;(\boldsymbol{N}(0),\boldsymbol{P}(0)))$ is the solution of \eqref{single species model}.
Then we have that either $\lim\limits_{t\rightarrow+\infty}N_i(t)=0$,
or $\lim\limits_{t\rightarrow+\infty}N_i(t)=\th_i$. Hence, our claim holds.

We next claim that $(\boldsymbol{0},\boldsymbol{0}),(\boldsymbol{\th},\boldsymbol{0})$ are uniform weak repellers in the sense that
\beq
\label{00 repeller}
\limsup\limits_{t\rightarrow+\infty}
\|\Theta(t)(\boldsymbol{U}(0),\boldsymbol{V}(0))-(\boldsymbol{0},\boldsymbol{0})\|\ge\delta_1\ \text{for all}\ (\boldsymbol{U}(0),\boldsymbol{V}(0))\in\p_0,
\eeq
and
\beq
\label{th0 repeller}
\limsup\limits_{t\rightarrow+\infty}
\|\Theta(t)(\boldsymbol{U}(0),\boldsymbol{V}(0))-(\boldsymbol{\th},\boldsymbol{0})\|\ge\delta_2\ \text{for all}\ (\boldsymbol{U}(0),\boldsymbol{V}(0))\in\p_0.
\eeq
Suppose by contradiction that \eqref{00 repeller} is not true.
Then for any $\delta>0$,
there exists $(\boldsymbol{U}(0),\boldsymbol{V}(0))\in\p_0$ such that
\beq
\limsup\limits_{t\rightarrow+\infty}
\|\Theta(t)(\boldsymbol{U}(0),\boldsymbol{V}(0))-(\boldsymbol{0},\boldsymbol{0})\|<\delta.
\eeq
Therefore, there exists $t_0>0$ such that
for $t\ge t_0$, we have that
\beq
\|\boldsymbol{N}(t,(\boldsymbol{N}(0),\boldsymbol{P}(0)))\|<\delta,
\|\boldsymbol{P}(t,(\boldsymbol{N}(0),\boldsymbol{P}(0)))\|<\delta.
\eeq
Consequently,
it follows from the equation for $\boldsymbol{N}$ that
\beq
\F{\d N_i}{\dt}\ge\sum\limits_{j=1}^{n}(d_1D_{ij}+qQ_{ij})N_j+N_i(r_i-\F{r_i\delta}{k_i}-a\delta),\ \ i\in\a,\ t\ge t_0.
\eeq
In view of $(\boldsymbol{U}(0),\boldsymbol{V}(0))\in\p_0$,
by Lemma \ref{Strong maximum principle},
we have $U_i(t_0)>0$, and then $N_i(t_0)>0$.
Thus there exists an $\al_0>0$ such that $N_i(t_0)\ge \al_0\phi_{1i}^{\del}$,
where $\boldsymbol{\phi}_{1}^{\del}$ is the principal
eigenfunction corresponding to the principal eigenvalue $\lam_1(d_1,q,\boldsymbol{r}-\F{\boldsymbol{r}}{\boldsymbol{k}}\del-\boldsymbol{a}\del)$ of \eqref{linear eigenvalue problem1}.
Let $\underline{N}_i(t)=\al_0 e^{\lam_1(d_1,q,\boldsymbol{r}-\f{\boldsymbol{r}}{\boldsymbol{k}}\del-\boldsymbol{a}\del)(t-t_0)}\phi_{1i}^{\delta}$.
Then $\underline{N}_i(t)$ satisfies
\beq
\left\{\arraycolsep=1.5pt
\begin{array}{ll}
\F{\d \underline{N}_i}{\dt}=\sum\limits_{j=1}^{n}(d_1D_{ij}+qQ_{ij})\underline{N}_j+\underline{N}_i(r_i-\F{r_i\delta}{k_i}-a\delta),\ \ &i\in\a,\ t>0,\\[2mm]
\boldsymbol{\underline{N}}(t_0)=\al_0\boldsymbol{\phi_1^{\del}}.&
\end{array}
\right.
\eeq
The comparison principle(Lemma \ref{Comparison principle}) yields that
\beq
N_i(t)\ge\underline{N}(t)\ \text{for all}\ t\ge t_0.
\eeq
Since $\lam_1(d_1,q,\boldsymbol{r})>0$,
by continuity,
we can choose $\delta>0$ small enough such that $\lam_1(d_1,q,\boldsymbol{r}-\F{\boldsymbol{r}}{\boldsymbol{k}}\del-\boldsymbol{a}\del)>0$,
which implies
$\lim\limits_{t\rightarrow+\infty}N_i(t,(\boldsymbol{N}(0),\boldsymbol{P}(0)))=+\infty$.
This is a contradiction to $\|\boldsymbol{N}(t,(\boldsymbol{N}(0),\boldsymbol{P}(0)))\|<\delta$ for $t>t_0$.
Hence, we conclude that $(\boldsymbol{0},\boldsymbol{0})$ is a uniform weak repeller.

Suppose by contradiction that \eqref{th0 repeller} is not true.
Then for any $\delta>0$,
there exists $(\boldsymbol{N}(0),\boldsymbol{P}(0))\in\p_0$ such that
$\limsup\limits_{t\rightarrow+\infty}\|\Theta(t)(\boldsymbol{N}(0),\boldsymbol{P}(0))-(\boldsymbol{\th},\boldsymbol{0})\|<\delta$. Therefore, there exists $t_1>0$ such that for $t\ge t_1$,
we have
\beq
\|\boldsymbol{N}(t,(\boldsymbol{N}(0),\boldsymbol{P}(0)))-\boldsymbol{\th}\|<\delta,
\|\boldsymbol{P}(t,(\boldsymbol{N}(0),\boldsymbol{P}(0)))\|<\delta.
\eeq
Therefore, it follows from the equation for $\boldsymbol{P}$ that
\beq
\F{\d P_i}{\dt}\ge\sum\limits_{j=1}^{n}(d_2D_{ij}+\tau q Q_{ij})P_j+P_i(\be a (\th_i-\delta)-\gam), \   \ i\in\a,\ t\ge t_1.
\eeq
In view of $(\boldsymbol{N}(0),\boldsymbol{P}(0))\in\p_0$,
by Lemma \ref{Strong maximum principle},
we have $P_i(t_1)>0$.
Let $\boldsymbol{\psi}_{1}^{\del}$ be the principal eigenfunction with respect to $\lam_1(d_2,\tau q,\be a(\boldsymbol{\th}-\boldsymbol{\delta})-\boldsymbol{\gam})$.
Thus
there exists an $\al_1>0$ such that $P_i(t_1)\ge \al_1\psi_{1i}^{\del}$. Let $\underline{P}_i(t)=\al_1 e^{\lam_1(d_2,\tau q,\be a(\boldsymbol{\th}-\boldsymbol{\delta})-\boldsymbol{\gam})(t-t_1)}\psi_{1i}^{\del}$.
Then $\underline{P}_i(t)$ satisfies
\beq
\left\{
\bea{l}
\F{\d \underline{P}_i}{\dt}=\sum\limits_{j=1}^{n}(d_2D_{ij}+\tau q Q_{ij})\underline{P}_j+\underline{P}_i(\be a (\th_i-\delta)-\gam),\ \ i\in\a,\ t\ge t_1,\\
\underline{P}_i(t_1)=\al_1\psi_{1i}^{\del}.
\eea
\right.
\eeq
It follows from Lemma \ref{Comparison principle} that
$P_i(t)\ge\underline{P}_i(t)$ for all $t\ge t_1$.
On the other hand, since $0<\gam<\gam^*$,
we have that $\lam_1(d_2,\tau q,\be a\boldsymbol{\th}-\boldsymbol{\gam})>0$ by \eqref{gam*}.
By continuity, we can choose $\delta>0$ small enough such that
$\lam_1(d_2,\tau q,\be a(\boldsymbol{\th}-\boldsymbol{\delta})-\boldsymbol{\gam})>0$,
which implies
$\lim\limits_{t\rightarrow+\infty}P_i(t,(\boldsymbol{N}(0),\boldsymbol{P}(0)))=+\infty$.
This is a contradiction to $\|\boldsymbol{P}(t,(\boldsymbol{N}(0),\boldsymbol{P}(0)))\|<\delta$ for
$t>t_1$.
Hence, we conclude that $(\boldsymbol{\th},\boldsymbol{0})$ is a uniform weak repeller.

We define now a continuous function $\mathcal{D}:\p\rightarrow[0,\infty)$ by
\beq
\mathcal{D}((\boldsymbol{N},\boldsymbol{P}))=\min\{\min\limits_{i\in\a}N_i,\min\limits_{i\in\a}P_i\}\ \text{for any}\ (\boldsymbol{N},\boldsymbol{P})\in\p.
\eeq
It follows from Lemma \ref{Strong maximum principle} that $\mathcal{D}^{-1}(0,\infty)\subset \p_0$ and $D$ satisfies
if $\mathcal{D}((\boldsymbol{N},\boldsymbol{P}))>0$ or $(\boldsymbol{N},\boldsymbol{P})\in\p_0$ with $\mathcal{D}((\boldsymbol{N},\boldsymbol{P}))=0$,
then $D(\Theta(t)(\boldsymbol{N},\boldsymbol{P}))>0$, $\forall~t>0$.
That is, $\mathcal{D}$ is a generalized distance function for the semiflow $\Theta(t):\p\rightarrow\p$ (see \cite{SZ}).
It follows from Step 2 that $\Theta(t)$ is point dissipative on $\p$.
Moreover,  it is
easy to see from the Arzela-Ascoli theorem that $\Theta(t):\p\rightarrow\p$ is compact for any $t>0$.
By \cite[Theorem 2.6]{MZ}, $\Theta(t):\p\rightarrow\p$, $t\ge 0$ admits a global compact attractor that attracts each bounded
set in $\p$.
Since $(\boldsymbol{0},\boldsymbol{0})$ and $(\boldsymbol{\th},0)$ are uniform weak repellers,
we conclude
that $(\boldsymbol{0},\boldsymbol{0})$ and $(\boldsymbol{\th},0)$ are isolated invariant sets in $\p$,
and
\beq
W^S\{(\boldsymbol{0},\boldsymbol{0})\}\cap\mathcal{D}^{-1}(0,\infty)=\emptyset,
W^S\{(\boldsymbol{\th},0)\}\cap\mathcal{D}^{-1}(0,\infty)=\emptyset,
\eeq
where $W^S\{(\boldsymbol{0},\boldsymbol{0})\},W^S\{(\boldsymbol{\th},0)\}$ are the stable sets of $(\boldsymbol{0},\boldsymbol{0}),(\boldsymbol{\th},0)$, respectively (see \cite{SZ}).
Hence,
there are no subsets of $\{(\boldsymbol{0},\boldsymbol{0})\}\cup\{(\boldsymbol{\th},0)\}$ form a cycle in $\pa\p_0$.
By \cite[Theorem 3]{SZ},
there exists $\eta>0$ such that for any $(\boldsymbol{N}(0),\boldsymbol{P}(0))\in\p_0$,
\beq
\min\limits_{(\boldsymbol{N},\boldsymbol{P})\in\omega((\boldsymbol{N}(0),\boldsymbol{P}(0)))}\mathcal{D}((\boldsymbol{N},\boldsymbol{P}))>\eta.
\eeq
This implies that for any $(\boldsymbol{N},\boldsymbol{P})\in\p_0$,
$\liminf\limits_{t\rightarrow+\infty}N_i(t)\ge\eta$ and $\liminf\limits_{t\rightarrow+\infty}P_i(t)\ge\eta$.
That is, system \eqref{predator prey patch} with initial conditions $(\boldsymbol{N}(0),\boldsymbol{P}(0))\in\p_0$ is
uniformly persistent.

\emph{Step 4.2. System \eqref{predator prey patch} has a steady state solution.}
It follows from \cite[Theorem 3.7 and Remark 3.10]{MZ} that
$\Theta(t):\p_0\rightarrow\p_0$ admits a global attractor $A_0$.
Then \cite[Theorem 4.7]{MZ} yields that $\Theta(t)$ admits at least one steady-state solution $(\boldsymbol{N}^*,\boldsymbol{P}^*)\in\p_0$.
Furthermore, we deduce that $\boldsymbol{N}^*,\boldsymbol{P}^*\gg\boldsymbol{0}$ by Lemma \ref{Strong maximum principle for elliptic}.
Thus, system \eqref{predator prey patch} admits at least one positive
steady state solution $(\boldsymbol{N}^*,\boldsymbol{P}^*)$.
That is, $(\boldsymbol{N}^*,\boldsymbol{P}^*)$ satisfies
\beq
\label{steady state}
\left\{\arraycolsep=1.5pt
\begin{array}{ll}
\sum\limits_{j=1}^{n}(d_1D_{ij}+qQ_{ij})N^*_j+r_iN^*_i(1-\F{N^*_i}{k_i})-aN^*_iP^*_i=0,\ \ &i\in\a,\\[2mm]
\sum\limits_{j=1}^{n}(d_2D_{ij}+\tau q Q_{ij})P^*_j+P^*_i(\be a N^*_i-\gam)=0, \   \ &i\in\a.
\end{array}
\right.
\eeq

The proof of this Theorem is hence completed.
\proofend

\ptb
(1) Note that $\gam^*(q,d_1,d_2)=\lam_1(d_2,\tau q,\be a\boldsymbol{\th})$.
By means of Lemma \ref{prop of lam1}(ii) and Lemma \ref{prop of th}(iv),
it follows that $\lim\limits_{q\rightarrow+\infty}\gam^*(q,d_1,d_2)=\be a k_n$.

(2) Since by Lemma \ref{prop of th}(v), there holds
$\lim\limits_{d_1\rightarrow+\infty}\th_i=\F{\sum_{i=1}^n r_i}{\sum_{i=1}^n\f{r_i}{k_i}}$,
it then indicates from Lemma \ref{hi=h0} that
\beq
\lim\limits_{d_1\rightarrow+\infty}\gam^*(q,d_1,d_2)=\F{\be a\sum_{i=1}^n r_i}{\sum_{i=1}^n\f{r_i}{k_i}}.
\eeq

(3)
Note from Lemma \ref{prop of th}(iv) that
$\lim\limits_{d_1\rightarrow0^+}(\th_1,\th_2,\cdots,\th_n)=(0,\cdots,0,k_n)$.
Hence one has $\lim\limits_{d_1\rightarrow0^+}\gam^*(q,d_1,d_2)=\mu(k_n)$,
where $\mu(k_n)$ is the principal eigenvalue of \eqref{mu(k_n)}.
Clearly, $\mu(0)=0$ and $\mu(k_n)$ is strictly increasing with respect to $k_n$.

(4)
Since $\gam^*(0,d_1,d_2)=\lam_1(d_2,0,\boldsymbol{\th}|_{q=0})$,
by Lemma \ref{prop of th}(vi),
there holds
\beq
\lim\limits_{d_1\rightarrow0^+}\gam^*(0,d_1,d_2)=\lam_1(d_2,0,\be a\boldsymbol{k}).
\eeq

(5) If $q\ge0$,
by applying Lemma \ref{prop of lam1}(iii),
one can have that
\beq
\lim\limits_{d_2\rightarrow0^+}\gam^*(q,d_1,d_2)
=\max\{\be a\th_1-\tau q,\cdots,\be a\th_{n-1}-\tau q,\be a\th_n\},
\ \text{and}\ \lim\limits_{d_2\rightarrow+\infty}\gam^*(q,d_1,d_2)
=\F{\sum_{i=1}^n\be a\th_i}{n}.
\eeq
The proof of
this Theorem is therefore completed.
\proofend

\subsection{Asymptotic profiles of positive steady states}

This subsection is devoted to studying the influence of diffusion and advection on the asymptotic profiles of the positive steady state of system \eqref{predator prey patch}.
Therefore, we focus on the steady states system \eqref{steady state}.
Let $U_i=(\F{d_1}{d_1+q})^{i-1}N_i$
and $V_i=(\F{d_2}{d_2+\tau q})^{i-1}P_i$.
Then $(\boldsymbol{U},\boldsymbol{V})$ satisfies
\beq
\label{elliptic UV}
\left\{\arraycolsep=1.5pt
\begin {array}{l}
-(d_1+q)U_1+(d_1+q)U_2+r_1U_1(1-\F{N_1}{k_1})-aU_1P_1=0,\\[2mm]
d_1 U_{j-1}+(-2d_1-q)U_{j}+(d_1+q)U_{j+1}+r_jU_j(1-\F{N_j}{k_j})-aU_jP_j=0,\ j\in\b,\\[2mm]
d_1U_{n-1}-d_1U_n+r_nU_n(1-\F{N_n}{k_n})-aU_nP_n=0,\\[2mm]
-(d_2+\tau q)V_1+(d_2+\tau q)V_2+V_1(\be a N_1-\gam)=0,\\[2mm]
d_2 V_{j-1}+(-2d_2-\tau q)V_{j}+(d_2+\tau q)V_{j+1}+V_j(\be a N_j-\gam)=0,\ j\in\b,\\[2mm]
d_2V_{n-1}-d_2V_n+V_n(\be a N_n-\gam)=0.
\end{array}\right.
\eeq

\begin{lem}
\label{N<th}
Suppose \eqref{assumption} hold.
Let $(\boldsymbol{N},\boldsymbol{P})$ be a nonnegative solution of \eqref{steady state} with $(\boldsymbol{N},\boldsymbol{P})\not\neq(\boldsymbol{0},\boldsymbol{0})$.
Then $\boldsymbol{0}\ll\boldsymbol{N}\ll\boldsymbol{\th}$ and $\gam<\gam^*(q,d_1,d_2)$.
\end{lem}

\proof
Note that, $(\boldsymbol{U},\boldsymbol{V})$ is a nonnegative solution of \eqref{elliptic UV} with $(\boldsymbol{U},\boldsymbol{V})\neq(\boldsymbol{0},\boldsymbol{0})$.
Lemma \ref{Strong maximum principle for elliptic} yields that $\boldsymbol{U}\gg\boldsymbol{0}$ and $\boldsymbol{V}\gg\boldsymbol{0}$.
Hence,
$\boldsymbol{N}\gg\boldsymbol{0}$ and $\boldsymbol{P}\gg\boldsymbol{0}$.
From the equations of $\boldsymbol{U}$ in \eqref{elliptic UV},
we have that
\beq
\left\{\arraycolsep=1.5pt
\begin {array}{l}
0=-(d_1+q)U_1+(d_1+q)U_2+r_1U_1(1-\F{N_1}{k_1})-aU_1P_1<-(d_1+q)U_1+(d_1+q)U_2+r_1U_1(1-\F{N_1}{k_1}),\\[2mm]
0=d_1 U_{j-1}+(-2d_1-q)U_{j}+(d_1+q)U_{j+1}+r_jU_j(1-\F{N_j}{k_j})-aU_jP_j\\[2mm]
<d_1 U_{j-1}+(-2d_1-q)U_{j}+(d_1+q)U_{j+1}+r_jU_j(1-\F{N_j}{k_j}),\ j\in\b,\\[2mm]
0=d_1U_{n-1}-d_1U_n+r_nU_n(1-\F{N_n}{k_n})<d_1U_{n-1}-d_1U_n+r_nU_n(1-\F{N_n}{k_n}).
\end{array}\right.
\eeq
Since $\boldsymbol{\th}$ is the unique positive solution to \eqref{single ss},
then $\boldsymbol{w}:=\big(\th_1,(\F{d_1}{d_1+q})\th_2,\cdots,(\F{d_1}{d_1+q})^{n-1}\th_n\big)$ is the unique positive solution to
\beq
\left\{\arraycolsep=1.5pt
\begin {array}{l}
(d_1+q)(w_2-w_1)+r_1(1-\F{w_1}{k_1})w_1=0,\\[2mm]
d_1w_{j-1}+(-2d_1-q)w_j+(d_1+q)w_{j+1}+r_j(1-(\F{d_1+q}{d_1})^{j-1}\F{w_j}{k_j})w_j=0,\ j\in\b,\\[2mm]
d_1(w_{n-1}-w_{n})+r_n(1-(\F{d_1+q}{d_1})^{n-1}\F{w_n}{k_n})w_n=0.
\end{array}\right.
\eeq
By the similar arguments in the proof of \cite[Lemma 2.2]{CLW2}(see also \cite{LT}),
one can conclude that $\boldsymbol{0}\ll\boldsymbol{U}\ll\boldsymbol{w}$,
which implies $\boldsymbol{0}\ll\boldsymbol{N}\ll\boldsymbol{\th}$.
Furthermore, from the equations of $\boldsymbol{V}$ in \eqref{elliptic UV}
and Lemma \ref{prop of lam1}(iv),
we have
\beq
\gam=\lam_1(d_2,\tau q,\be a\boldsymbol{N})<\lam_1(d_2,\tau q,\be a\boldsymbol{\th})=\gam^*(q,d_1,d_2).
\eeq
This completes the proof.
\proofend

\begin{lem}
\label{increasing NP}
Suppose \eqref{assumption} hold.
Let $(\boldsymbol{N},\boldsymbol{P})$ be a nonnegative solution of \eqref{steady state} with $(\boldsymbol{N},\boldsymbol{P})\neq(\boldsymbol{0},\boldsymbol{0})$.
We have the following results:

(1) if $q\ge r_{max}+2\sqrt{d_1 r_{max}}$,
there holds $0<N_1<\cdots<N_n$;

(2) if $q\ge\max\big\{r_{max}+2\sqrt{d_1 r_{max}},\F{\be aN_n+2\sqrt{\be a d_2N_n}}{\tau}\big\}$,
there holds $0<P_1<\cdots<P_n$.
\end{lem}

\proof
(1) It follows from Lemma \ref{N<th} that $\boldsymbol{N}\gg\boldsymbol{0}$ and $\boldsymbol{P}\gg\boldsymbol{0}$.
Now set $\al_i=(\F{d_1}{d_1+q})^{\f{i-1}{2}}$,
$\xi_i=\al_iN_i$
and $\zeta_i=\al_iP_i$.
Direct calculations yield
\beq
\label{08}
\left\{\arraycolsep=1.5pt
\begin {array}{l}
\sqrt{d_1(d_1+q)}(\xi_2-\xi_1)+\big[r_1(1-\F{N_1}{k_1})-d_1-q+\sqrt{d_1(d_1+q)}-\F{a\zeta_1}{\al_1}\big]\xi_1=0,\\[2mm]
\sqrt{d_1(d_1+q)}(\xi_{j-1}-2\xi_j+\xi_{j+1})+\big[r_j(1-\F{N_j}{k_j})-2d_1-q+2\sqrt{d_1(d_1+q)}-\F{a\zeta_j}{\al_j}\big]\xi_j=0,\ j\in\b.
\end{array}
\right.
\eeq
If $q\ge r_{max}+2\sqrt{d_1r_{max}}$, then
\beq
\left\{\arraycolsep=1.5pt
\begin {array}{l}
r_1(1-\F{N_1}{k_1})-d_1-q+\sqrt{d_1(d_1+q)}-\F{a\zeta_1}{\al_1}<0,\\[2mm]
r_j(1-\F{N_j}{k_j})-2d_1-q+\sqrt{d_1(d_1+q)}-\F{a\zeta_j}{\al_j}<0,\ j\in\b.
\end{array}
\right.
\eeq
Thus, we get from \eqref{08} that $0<\xi_1<\xi_2<\cdots<\xi_n$.
This indicates that $0<N_1<\cdots<N_n$.

(2) Set $a_i=(\F{d_2}{d_2+\tau q})^{\f{i-1}{2}}$, $\xi_i=a_iN_i$ and $\zeta_i=a_iP_i$.
Then we have
\beq
\label{09}
\left\{\arraycolsep=1.5pt
\begin {array}{l}
\sqrt{d_2(d_2+\tau q)}(\zeta_2-\zeta_1)+\big[\be a N_1-\gam-d_2-\tau q+\sqrt{d_2(d_2+\tau q)}\big]\zeta_1=0,\\[2mm]
\sqrt{d_2(d_2+\tau q)}(\zeta_{j-1}-2\zeta_j+\zeta_{j+1})+\big[\be a N_j-\gam-2d_2-\tau q+\sqrt{d_2(d_2+\tau q)}\big]\zeta_j=0,\ j\in\b.
\end{array}
\right.
\eeq
By means of (1),
we get that
\beq
\left\{\arraycolsep=1.5pt
\begin {array}{l}
\be a N_1-\gam-d_2-\tau q+\sqrt{d_2(d_2+\tau q)}
<\be a N_n-d_2-\tau q+\sqrt{d_2(d_2+\tau q)}\le0,\\[2mm]
\be a N_i-\gam-2d_2-\tau q+\sqrt{d_2(d_2+\tau q)}
<\be a N_n-2d_2-\tau q+\sqrt{d_2(d_2+\tau q)}\le0,\ j\in\b.
\end{array}
\right.
\eeq
Here
$q\ge\max\big\{r_{max}+2\sqrt{d_1 r_{max}},\F{\be aN_n+2\sqrt{\be a d_2N_n}}{\tau}\big\}$
is used.
Hence, we get from \eqref{09} that $0<\zeta_1<\zeta_2<\cdots<\zeta_n$.
This indicates that $0<P_1<\cdots<P_n$.
The proof of
this lemma is therefore completed.
\proofend

\begin{lem}
\label{Pi bounded}
Suppose \eqref{assumption} hold.
Let $(\boldsymbol{N},\boldsymbol{P})$ be a nonnegative solution of \eqref{steady state} with $(\boldsymbol{N},\boldsymbol{P})\neq(\boldsymbol{0},\boldsymbol{0})$.
Then the following results hold:

(1) for fixed $d_1,d_2>0$,
$P_i$ is uniformly bounded for $i\in\a$ as $q\rightarrow+\infty$;

%(2) for fixed $d_2>0$ and $q>2\overline{r}$,
%$P_i$ is uniformly bounded for $i\in\a$ as $d_1\rightarrow0^+$;

(2) for fixed $d_1>0$ and
$q>\max\big\{r_{max}+2\sqrt{d_1 r_{max}},\F{\be aN_n}{\tau}\big\}$,
$P_i$ is uniformly bounded for $i\in\a$ as $d_2\rightarrow0^+$;

(3) for fixed $d_2>0$ and $q\ge0$,
$P_i$ is uniformly bounded for $i\in\a$ as $d_1\rightarrow+\infty$;

(4) for fixed $d_1>0$ and $q\ge0$,
$P_i$ is uniformly bounded for $i\in\a$ as $d_2\rightarrow+\infty$.
\end{lem}

\proof
By a similar argument as in the proof of Lemma \ref{N<th},
we have $\boldsymbol{N}\gg\boldsymbol{0}$ and $\boldsymbol{P}\gg\boldsymbol{0}$.

(1)
According to Lemma \ref{increasing NP}(2),
we only need to show $P_n$ is uniformly bounded as $q\rightarrow+\infty$.
If this is not true, then by passing to a sequence if necessary, we may assume $P_n\rightarrow+\infty$ as $q\rightarrow+\infty$.
Set $\tilde{P}_i=\F{P_i}{P_n}$.
Then
\beq
\label{equations of N tildeP}
\left\{\arraycolsep=1.5pt
\begin{array}{ll}
\sum\limits_{j=1}^{n}(d_1D_{ij}+qQ_{ij})N_j+r_iN_i(1-\F{N_i}{k_i})-aN_i\tilde{P}_iP_n=0,\ \ &i\in\a,\\[2mm]
\sum\limits_{j=1}^{n}(d_2D_{ij}+\tau q Q_{ij})\tilde{P}_j+\tilde{P}_i(\be a N_i-\gam)=0, \   \ &i\in\a.
\end{array}
\right.
\eeq

Since $k_i\le k_{max}$, $0<\tilde{P}_i\le1$ and $0<N_i<\th_i$ is bounded uniformly for $q\rightarrow+\infty$ by lemma \ref{prop of th} and Lemma \ref{N<th},
passing to a sequence if necessary yields that
$k_i\rightarrow k_i^*\ge0$,
$\tilde{P}_i\rightarrow P^*_i\le1$ and
$N_i\rightarrow N^*_i\ge0$.
In particular,
$P^*_n=1$, since $\tilde{P}_n=1$.
By dividing the second equation of \eqref{equations of N tildeP} by $q$,
it follows that
\beq
\left\{\arraycolsep=1.5pt
\begin{array}{ll}
(-\F{d_2}{q}-\tau)\tilde{P}_1+\F{d_2}{q}\tilde{P}_2+\F{\be a N_1-\gam}{q}\tilde{P}_1=0,&\\[2mm]
(\F{d_2}{q}+\tau)\tilde{P}_{j-1}+(-\F{2d_2}{q}-\tau)\tilde{P}_j+\F{d_2}{q}\tilde{P}_{j+1}+\F{\be aN_j-\gam}{q}\tilde{P}_j=0,&j\in\b.
\end{array}
\right.
\eeq
Consequently, as $q\rightarrow+\infty$,
$\tilde{P}_1,\tilde{P}_2,\cdots,\tilde{P}_{n-1}\rightarrow0$.
Noting that
$\sum\limits_{i=1}^{n}\sum\limits_{j=1}^{n}(d_2D_{ij}+\tau qQ_{ij})\tilde{P}_j=0$,
we deduce that
\beq
\sum\limits_{i=1}^{n}\tilde{P}_i(\be aN_i-\gam)=0.
\eeq
A simple calculation gives $\be aN_n^*-\gam=0$,
which shows $N_n^*=\F{\gam}{\be a}$.

Combining the
first equation of \eqref{equations of N tildeP} with $\sum\limits_{i=1}^{n}\sum\limits_{j=1}^{n}(d_1D_{ij}+qQ_{ij})N_j=0$ yields that
\beq aP_n\sum\limits_{i=1}^{n}N_i\tilde{P}_i=\sum\limits_{i=1}^{n}N_ir_i(1-\F{N_i}{k_i}),
\eeq
which implies that
\beq
\label{11}
+\infty\leftarrow aP_n=\F{\sum\limits_{i=1}^{n}N_ir_i(1-\F{N_i}{k_i})}{\sum\limits_{i=1}^{n}N_i\tilde{P}_i}.
\eeq
This is impossible.
Hence $P_n$ is uniformly bounded as $q\rightarrow+\infty$.

(2) Given $q>\max\big\{r_{max}+2\sqrt{d_1 r_{max}},\F{\be aN_n}{\tau}\big\}$,
by the similar arguments as in the proof of Lemma \ref{increasing NP},
we can deduce that for
$d_2\ll 1$,
$0<P_1<\cdots<P_n$.
Hence, we need only to show $P_n$ is bounded uniformly for $d_2\rightarrow0^+$.
Suppose that there exist a sequence $d_2^m$ and positive solutions $(\boldsymbol{N}^m,\boldsymbol{P}^m)$ to \eqref{steady state} such that $\lim\limits_{m\rightarrow+\infty}d_2^m=0$ and $P_n^m\rightarrow+\infty$ as $m\rightarrow+\infty$.
Set $\tilde{P}_i^m=\F{P_i^m}{P_n^m}\le1$.
Then we reach that
\beq
\left
\{\arraycolsep=1.5pt
\begin{array}{ll}
(-d_2^m-\tau q)\tilde{P}^m_1+d_2\tilde{P}^m_2+(\be a N_1-\gam)\tilde{P}^m_1=0,&\\[2mm]
(d_2^m+\tau q)\tilde{P}^m_{j-1}+(-2d_2^m-\tau q)\tilde{P}^m_j+d_2^m\tilde{P}^m_{j+1}+(\be aN_j-\gam)\tilde{P}^m_j=0,&j\in\b,\\[2mm]
(d_2^m+\tau q)\tilde{P}^m_{n-1}-d_2^m\tilde{P}^m_n+(\be a N_n-\gam)\tilde{P}^m_n=0.
\end{array}
\right.
\eeq
Since $N_i$ is bounded for all $d_2>0$,
we may assume that
$\tilde{P}_i^m\rightarrow P^*_i\in[0,1]$ and $N_i^m\rightarrow N^*_i\ge0$
as $m\rightarrow+\infty$.
Letting $m\rightarrow+\infty$
yields
\beq
\left
\{\arraycolsep=1.5pt
\begin{array}{ll}
-\tau qP^*_1+(\be a N_1^*-\gam)P^*_1=0,&\\[2mm]
\tau qP^*_{j-1}-\tau qP^*_j+(\be aN_j^*-\gam)P^*_j=0,&j\in\b,\\[2mm]
\tau qP^*_{n-1}+(\be a N_n^*-\gam)P^*_n=0,
\end{array}
\right.
\eeq
from which
$P^*_1=0$ or $\be a N_1^*=\gam+\tau q>0$ follows.
On the other hand,
by dividing the equation of $N_i^m$ by $P_n^m$ and letting $m\rightarrow+\infty$,
one can deduce that $N^*_1P^*_1=0$.
Hence, $P^*_1=0$.
By the similar methods above,
it further follows that
$P^*_1=\cdots=P^*_{n-1}=0$,
$P^*_n=1$.
Therefore, a contradiction happens according to the same arguments of deriving \eqref{11}.
Hence, $P_n$ is bounded for $d_2\rightarrow0^+$.

(3) Note that $N_i$ is bounded uniformly for $d_1\rightarrow+\infty$.
Suppose that $P_i$ is unbounded for $i\in\a$ as $d_1\rightarrow+\infty$.
There exist a sequence $d_1^m$ and positive solutions $(\boldsymbol{N}^m,\boldsymbol{P}^m)$ to \eqref{steady state} such that
$d_1^m\rightarrow+\infty$ and
$P_{i_0}^m:=\max\limits_{i\in\a}P_i^m\rightarrow+\infty$ as $m\rightarrow+\infty$.
Let $\hat{P}_i^m=\F{P_{i}^m}{P_{i_0}^m}\le1$.
Then we can deduce that there exists a convergent subsequence of ($N_i^m$,$\hat{P}_i^m$)(still denoted by ($N_i^m$,$\hat{P}_i^m$))
such that $\hat{P}_i^m\rightarrow\hat{P}_i\in[0,1]$,
$N_i^m\rightarrow N_i\ge0$ as $m\rightarrow+\infty$.
Clearly, $1=\hat{P}_{i_0}^m\rightarrow\hat{P}_{i_0}>0$ as $m\rightarrow+\infty$.

Similar to the proof of (2),
we also have $\boldsymbol{\hat{P}}>\boldsymbol{0}$ and
\beq
\lam_1(d^m_1,q,r_i(1-\F{N_i^m}{k_i})-a\hat{P}_i^mP^m_{i_0})=0.
\eeq
By Lemma \ref{prop of lam1}(iii) again,
we have $\lam_1(d^m_1,q,r_i(1-\F{N_i^m}{k_i})-a\hat{P}_i^mP^m_{i_0})<0$ as $m\rightarrow+\infty$,
a contradiction occurs.
Therefore, $P_i$ is bounded for $i\in\a$ as $d_1\rightarrow+\infty$.

(4) Suppose now that $P_i$ is unbounded for $i\in\a$ as $d_2\rightarrow+\infty$.
There exist a sequence $d_2^m$ and positive solutions $(\boldsymbol{N}^m,\boldsymbol{P}^m)$ to \eqref{steady state} such that
$d_2^m\rightarrow+\infty$ and
$P_{i_0}^m:=\max\limits_{i\in\a}P_i^m\rightarrow+\infty$ as $m\rightarrow+\infty$.
Let $\hat{P}_i^m=\F{P_{i}^m}{P_{i_0}^m}\le1$.
Then we can deduce that there exists a convergent subsequence of ($N_i^m$,$\hat{P}_i^m$)(still denoted by ($N_i^m$,$\hat{P}_i^m$))
satisfying $\hat{P}_i^m\rightarrow\hat{P}_i\in[0,1]$,
$N_i^m\rightarrow N_i\ge0$ as $m\rightarrow+\infty$.
Clearly, $1=\hat{P}_{i_0}^m\rightarrow\hat{P}_{i_0}>0$ as $m\rightarrow+\infty$.

Dividing the equation of $\hat{P}_i^m$ by $d_2^m$ and letting $m\rightarrow+\infty$ yield that $\hat{P}_1=\cdots=\hat{P}_n=\hat{P}_{i_0}=1$.
Therefore, a contradictions occurs according the same arguments of deriving \eqref{11}.
$P_i$ is hence bounded for $d_2\rightarrow+\infty$.

The proof of this lemma is thus complete.
\proofend

It follows from Theorem \ref{critical curve}(2) that
\eqref{predator prey patch} has a positive steady state $(\boldsymbol{N}^*,\boldsymbol{P}^*)$ when $0<\gam<\gam^*(q,d_1,d_2)$.
In the following,
we shall investigate the positive steady state $(\boldsymbol{N}^*,\boldsymbol{P}^*)$.

\begin{lem}
Suppose \eqref{assumption} hold,
$0<\gam<\gam^*(q,d_1,d_2)$.
Then for $i\in\c$,
\beq
\label{estimate UV}
\left\{
\bea{l}
(\F{d_1+q+2(aP^*_n+\max_{i\in\a}\f{r_i}{k_i}N^*_n)}{d_1})^{i-n}N^*_n<N^*_i<(\F{d_1+q-2r_{max}}{d_1})^{i-n}N^*_n,\\
\text{if}\ q\ge\max\{d_1+2r_{max},r_{max}+2\sqrt{d_1r_{max}}\},\\
(\F{d_2+\tau q+2\gam}{d_2})^{i-n}P^*_n<P^*_i<(\F{d_2+\tau q-2\be aN^*_n}{d_2})^{i-n}P^*_n,\ \text{if}\ q\ge\max\{\F{d_2+2\be aN^*_n}{\tau},\F{\be aN^*_n+2\sqrt{\be aN^*_nd_2}}{\tau}\}.
%,\F{d_2-2\gam}{\tau}\}.
\eea
\right.
\eeq
\end{lem}

\proof
Set $\al=\F{d_1+q-2r_{max}}{d_1}$,
$\del=\F{d_1+q+2(aP^*_n+\max_{i\in\a}\f{r_i}{k_i}N^*_n)}{d_1}$,
$A=\F{d_2+\tau q-2\be aN^*_n}{d_2}$,
$B=\F{d_2+\tau q+2\gam}{d_2}$,
$\al_i=\al^{i-1}$,
$\del_i=\del^{i-1}$,
$A_i=A^{i-1}$, $B_i=B^{i-1}$.
Let $\overline{N}^*_i=N_n^*\F{\al_i}{\al_n}$,
$\underline{N}_i^*=N_n^*\F{\del_i}{\del_n}$,
$\overline{P}_i^*=P_n^*\F{A_i}{A_n}$,
$\underline{P}_i^*=P_n^*\F{B_i}{B_n}$.
Letting $u_i=(\F{d_1}{d_1+q})^{\f{i-1}{2}}(\overline{N}^*_i-N^*_i)$,
it is obvious that
\beq
\left\{\arraycolsep=1.5pt
\begin {array}{l}
\sqrt{d_1(d_1+q)}(u_2-u_1)+\big[r_1(1-\F{N^*_1}{k_1})-d_1-q+\sqrt{d_1(d_1+q)}-aP^*_1\big]u_1<0,\\[2mm]
\sqrt{d_1(d_1+q)}(u_{j-1}-2u_j+u_{j+1})+\big[r_j(1-\F{N^*_j}{k_j})-2d_1-q+2\sqrt{d_1(d_1+q)}-aP^*_j\big]u_j<0,\ j\in\b,
\end{array}
\right.
\eeq
for $q\ge\max\{d_1+2r_{max},r_{max}+2\sqrt{d_1r_{max}}\}$,
some similar methods in obtaining \eqref{inequality of thi} give rise to $u_i>0$ and thus $\overline{N}^*_i>N^*_i$ for $i\in\c$.

By means of the similar arguments above,
combining the equations of $\underline{N}_i^*,\overline{P}_i^*,\underline{P}_i^*$ with
\beq
q\ge\max\{d_1,\F{d_2+2\be ak_n}{\tau},\F{\be ak_n+2\sqrt{\be ak_nd_2}}{\tau},\F{d_2-2\gam}{\tau}\}
\eeq
yields that $N^*_i>\underline{N}^*_i,\overline{P}^*_i>P^*_i>\underline{P}^*_i$ for $i\in\c$.
Hence, we finish the proof of this lemma.
\proofend

\ptc
Note that $0<\gam<\be ak_n$ and $\lim\limits_{q\rightarrow+\infty}\gam^*(q,d_1,d_2)=\be a k_n$
guarantee that
there exists a sufficiently large positive constant $Q$ such that $0<\gam<\gam^*(q,d_1,d_2)$ when $q>Q$.
This implies that
a positive steady state of system \eqref{predator prey patch} exists for all large $q$(see Theorem \ref{critical curve}(2)).

Notice from Lemmas \ref{N<th} and \ref{Pi bounded}(1) that
$N^*_n$ and $P^*_n$ are uniformly bounded as $q\rightarrow+\infty$.
By choosing a subsequence if necessary,
we may assume that
$(N^*_n,P^*_n)\rightarrow(p_1,p_2)$
as $q\rightarrow+\infty$ for some constants $p_1$ and $p_2$.
It follows from \eqref{estimate UV} that
\beq
\F{N^*_i}{N^*_n}\rightarrow0,
\F{P^*_i}{P^*_n}\rightarrow0,\ \text{for}\ i\in\c, q\rightarrow+\infty.
\eeq
Applying \eqref{steady state},
we get
\beq
\left\{\arraycolsep=1.5pt
\begin{array}{l}
\sum\limits_{i=1}^{n}\F{N^*_i}{N^*_n}(r_i-\F{r_i}{k_i}\F{N^*_i}{N^*_n}N^*_n-a\F{P^*_i}{P^*_n}P^*_n)=0,\\[2mm]
\sum\limits_{i=1}^{n}\F{P^*_i}{P^*_n}(\be a \F{N^*_i}{N^*_n}N^*_n-\gam)=0.
\end{array}
\right.
\eeq
Furthermore,
we reach that
\beq
\left\{\arraycolsep=1.5pt
\begin{array}{l}
0=r_n-\F{r_n}{k_n}N^*_n-aP^*_n+\sum\limits_{i=1}^{n-1}\F{N^*_i}{N^*_n}(r_i-\F{r_i}{k_i}\F{N^*_i}{N^*_n}N^*_n-a\F{P^*_i}{P^*_n}P^*_n)\rightarrow r_n-\F{r_n}{k_n}p_1-ap_2,\\[2mm]
0=\be a N^*_n-\gam+\sum\limits_{i=1}^{n-1}\F{P^*_i}{P^*_n}(\be a \F{N^*_i}{N^*_n}N^*_n-\gam)\rightarrow\be a p_1-\gam,
\end{array}
\right.
\eeq
by letting
$q\rightarrow+\infty$.
Therefore,
we obtain that
\beq
(p_1,p_2)=(\F{\gam}{\be a},\F{1}{a}(r_n-\F{r_n}{k_n}\F{\gam}{\be a})).
\eeq
According to \eqref{estimate UV},
we have
\beq
\bea{l}
\big[(\F{d_1}{d_1+q+2(aP^*_n+\max_{i\in\a}\f{r_i}{k_i}N^*_n)})^{n-i}-(\F{d_1}{d_1+q})^{n-i}\big]N^*_n<N^*_i-N^*_n(\F{d_1}{d_1+q})^{n-i}\\
<\big[(\F{d_1}{d_1+q-2r_{max}})^{n-i}-(\F{d_1}{d_1+q})^{n-i}\big]N^*_n,
\eea
\eeq
which implies
\beq
\max\limits_{i\in\a}\Big(N^*_i-N^*_n(\F{d_1}{d_1+q})^{n-i}\Big)\rightarrow0\ \text{as}\ q\rightarrow+\infty.
\eeq
Similarly,
it can be proved by \eqref{estimate UV} that
\beq
\max\limits_{i\in\a}\Big(P^*_i-P^*_n(\F{d_2}{d_2+\tau q})^{n-i}\Big)\rightarrow0\ \text{as}\ q\rightarrow+\infty.
\eeq
That is,
\beq
\max\{\max\limits_{i\in\a}\Big(N^*_i-N^*_n(\F{d_1}{d_1+q})^{n-i}\Big),\max\limits_{i\in\a}\Big(P^*_i-P^*_n(\F{d_2}{d_2+\tau q})^{n-i}\Big)\}\rightarrow0\ \text{as}\ q\rightarrow+\infty,
\eeq
which implies that $(N^*_i,P^*_i)\rightarrow(0,0)$ uniformly for $i\in\c$ as $q\rightarrow+\infty$.
In conclusion,
we complete the proof of Theorem \ref{UV q infty}.
\proofend

\ptd
Note that $0<\gam<\be ak_{min}$ and $\lim\limits_{d_1\rightarrow+\infty}\gam^*(q,d_1,d_2)=\F{\be a\sum_{i=1}^n r_i}{\sum_{i=1}^n\f{r_i}{k_i}}>\be ak_{min}$
guarantee that
there exists a sufficiently large positive constant $D_1$ such that $0<\gam<\gam^*(q,d_1,d_2)$ when $d_1>D_1$.
This implies that
a positive steady state of system \eqref{predator prey patch} exists for all large $d_1$(see Theorem \ref{critical curve}(2)).

It follows from Lemmas \ref{N<th}, \ref{prop of th}(v) and \ref{Pi bounded}(3) that
$N^*_i$ and $P^*_i$ are uniformly bounded for $i\in\a$ as $d_1\rightarrow+\infty$.
We may assume by passing to a subsequence if necessary that
$P^*_i\rightarrow \dot{P}_i$ as $d_1\rightarrow+\infty$.
Dividing the equation of $N^*_i$ in \eqref{steady state} by $d_1$ and taking $d_1\rightarrow+\infty$
lead to $N^*_i\rightarrow s\ge0$ for all $i$.
Let $\max\limits_{i\in\a}P^*_i=P^*_{i_0(d_1)}$ and $\tilde{P}_i:=\F{P^*_i}{P^*_{i_0(d_1)}}$.
Then it follows that
$\tilde{P}_i\rightarrow\F{\dot{P}_i}{P^*_{i_0(\infty)}}\ge0$ for $d_1\rightarrow+\infty$.
Clearly, $\tilde{P}_i$ satisfies
\beq
\sum\limits_{j=1}^{n}(d_2D_{ij}+\tau q Q_{ij})\tilde{P}_j+\tilde{P}_i(\be a N^*_i-\gam)=0.
\eeq
After taking $d_1\rightarrow+\infty$,
we arrive at
\beq
\label{16}
\sum\limits_{j=1}^{n}(d_2D_{ij}+\tau q Q_{ij})\F{\dot{P}_j}{P^*_{i_0(\infty)}}+\F{\dot{P}_i}{P^*_{i_0(\infty)}}(\be a s-\gam)=0.
\eeq
Adding \eqref{16} up with respect to $i$
yields
\beq
\sum\limits_{i=1}^n\F{\dot{P}_i}{P^*_{i_0(\infty)}}(\be a s-\gam)=0.
\eeq
The above equality then implies that $s=\F{\gam}{\be a}$.
Letting $d_1\rightarrow+\infty$ in the equation of $P^*_i$ in \eqref{steady state},
we see that
\beq
\sum\limits_{j=1}^{n}(d_2D_{ij}+\tau q Q_{ij})\dot{P}_j=0,
\eeq
which implies that $\dot{P}_i=(1+\F{\tau q}{d_2})^{i-1}\dot{P}_1$.

After
letting $d_1\rightarrow+\infty$ in the equation of $N^*_i$ in \eqref{steady state}
and adding them up respect to $i$ from $1$ to $n$,
we can
see that
\beq
\sum\limits_{i=1}^nr_i(1-\F{\gam}{\be ak_i})=a\dot{P}_1\sum\limits_{i=1}^n(1+\F{\tau q}{d_2})^{i-1}.
\eeq
That is,
$\dot{P}_1=(1+\F{\tau q}{d_2})^{i-1}\F{\sum_{j=1}^nr_j(1-\f{\gam}{\be ak_j})}{\f{d_2a}{\tau q}[(1+\f{\tau q}{d_2})^n-1]})$.
Then \eqref{12} is obtained.
\proofend

\pte
Note that $0<\gam<\gam^*(q,d_1,d_2)$.
This implies that
a positive steady state of system \eqref{predator prey patch} exists(see Theorem \ref{critical curve}(2)).

Note that $N^*_i$ is uniformly bounded as $d_2\rightarrow0^+$(see Lemma \ref{N<th}).
Since
$q>\max\big\{r_{max}+2\sqrt{d_1 r_{max}},\F{2\be aN^*_n}{\tau}\}$,
by Lemma \ref{Pi bounded}(2) and \eqref{estimate UV},
we have that $P^*_n$ is uniformly bounded and
\beq
(\F{d_2+\tau q+2\gam}{d_2})^{i-n}P^*_n<P^*_i<(\F{d_2+\tau q-2\be aN^*_n}{d_2})^{i-n}P^*_n
\eeq
for $d_2\rightarrow0^+$.
Therefore,
\beq
\max\limits_{i}\Big(P^*_i-P^*_n(\F{d_1}{d_2+\tau q})^{n-i}\Big)\rightarrow0\ \text{as}\ d_2\rightarrow0^+,
\eeq
and we may suppose $(N^*_i,P^*_n)\rightarrow(N^0_i,l)$ as $d_2\rightarrow0^+$ by passing to a subsequence if necessary.
Here $N^0_i\ge0,l\ge0$.

Set $\tilde{P}_i=\F{P^*_i}{P^*_n}$.
Then $\tilde{P}_i$ satisfies
\beq
\label{13}
\sum\limits_{j=1}^{n}(d_2D_{ij}+\tau q Q_{ij})\tilde{P}_j+\tilde{P}_i(\be a N^*_i-\gam)=0.
\eeq
If $l>0$, then we must have $\tilde{P}_i\rightarrow0$($i\in\c$) as $d_2\rightarrow0^+$.
Combined with \eqref{13},
we obtain that $l=0$.
Setting $d_2\rightarrow0^+$ in the equation of $N^*_i$ yields $N^0_i=\th_i$.
The proof of Theorem \ref{UV d2 0} is thus complete.
\proofend

At last, before giving the proof of Theorem \ref{UV d2 infty},
we introduce the following lemma.

\begin{lem}
\label{Z}
Suppose \eqref{assumption} hold, $0<\gam<\F{\be a\sum_i^n\th_i}{n}$
and $y\ge0$.
The system
\beq
\label{14}
\left\{\arraycolsep=1.5pt
\begin{array}{ll}
\sum\limits_{j=1}^{n}(d_1D_{ij}+qQ_{ij})Z_j+Z_i(r_i-\F{r_i}{k_i}Z_i-ay)=0,\ \ &i\in\a,\\[2mm]
\be a\sum\limits_{i=1}^{n}Z_i=\gam n, \   \ &i\in\a.
\end{array}
\right.
\eeq
admits a unique positive solution $(Z_i(\tilde{y}),\tilde{y})$.
\end{lem}

\proof
Let $y^*=\F{\lam_1(d_1,q,\boldsymbol{r})}{a}>0$.
Clearly, it follows from \cite[Lemma 2.2]{CLW2}(see also \cite{LT}) that
for $y\ge y^*$,
$\boldsymbol{Z}=\boldsymbol{0}$ is the unique solution of the first equation of \eqref{14},
and for $0\le y<y^*$, the first equation of \eqref{14} admits a unique positive solution,
denoted by $Z_i(y)$.
Now choosing $0\le y^1<y^2<y^*$,
we
obtain
\beq
\sum\limits_{j=1}^{n}(d_1D_{ij}+qQ_{ij})Z_j+Z_i(r_i-\F{r_i}{k_i}Z_i-ay^2)<
\sum\limits_{j=1}^{n}(d_1D_{ij}+qQ_{ij})Z_j+Z_i(r_i-\F{r_i}{k_i}Z_i-ay^1).
\eeq
By applying some similar arguments in the proof of \cite[Lemma 2.2]{CLW2}(see also \cite{LT}),
we have $0<Z_i(y^2)<Z_i(y^1)$
which
yields $Z_i(y)$ is strictly decreasing with respect to $y$ in $[0,y^*]$.
Thus, $F(y)=\sum\limits_{i=1}^nZ_i(y)-\F{\gam n}{\be a}$ is strictly decreasing with respect to $y$ in $[0,y^*]$.
Since $F(0)=\sum\limits_{i=1}^nZ_i(0)-\F{\gam n}{\be a}=\sum\limits_{i=1}^n\th_i-\F{\gam n}{\be a}>0$,
$F(y^*)=-\F{\gam n}{\be a}<0$,
we can deduce from the monotonicity of $F(y)$ that
there exists a unique $\tilde{y}\in(0,y^*)$
such that $F(\tilde{y})=0$.
After substituting $\tilde{y}$ into the first equation of \eqref{14},
we deduce that \eqref{14} admits a unique positive solution $(Z_i(\tilde{y}),\tilde{y})$.
\proofend

\ptf
Note that $0<\gam<\F{\be a\sum_i^n\th_i}{n}$ and $\lim\limits_{d_2\rightarrow+\infty}\gam^*(q,d_1,d_2)=\F{\be a\sum_i^n\th_i}{n}$
guarantee that
there exists a sufficiently large positive constant $D_2$ such that $0<\gam<\gam^*(q,d_1,d_2)$ when $d_2>D_2$.
This implies that
a positive steady state of system \eqref{predator prey patch} exists for all large $d_2$(see Theorem \ref{critical curve}(2)).

It can be shown similarly to Lemma \ref{N<th} and \ref{Pi bounded}(4) that
$N^*_i,P^*_i$ are uniformly bounded as $d_2\rightarrow+\infty$.
By using similar arguments as in the proof of Theorem \ref{UV d1 infty},
we may assume that
$(N^*_i,P^*_i)\rightarrow(\hat{\th}_i,z)$ as $d_2\rightarrow+\infty$ by passing to a subsequence if necessary,
where $(\hat{\th}_i,z)$ satisfies
\beq
\left\{\arraycolsep=1.5pt
\begin{array}{ll}
\sum\limits_{j=1}^{n}(d_1D_{ij}+qQ_{ij})\hat{\th}_j+\hat{\th}_i(r_i-\F{r_i}{k_i}\hat{\th}_i-az)=0,\ \ &i\in\a,\\[2mm]
\sum\limits_{i=1}^{n}(\be a \hat{\th}_i-\gam)=0, \   \ &i\in\a.
\end{array}
\right.
\eeq
By Lemma \ref{Z},
it follows that $(\hat{\th}_i,z)$ exists and is unique.
\proofend

\section{Discussion}

In this paper,
we consider the specialist predator-prey model \eqref{predator prey patch} in a closed advective patchy environment.
We obtain threshold dynamics in terms of the mortality rate of the specialist predators.
In Theorem \ref{critical curve},
we prove that the specialist predators will die out if their mortality rate exceeds a critical value $\gam^*$ and persist if the mortality rate is less than $\gam^*$.
%In Theorems \ref{properties of critical curve},
%we show that advection and diffusion have little effects on the invasion of predators
%unless the prey's diffusion rate is sufficiently small.
In Theorem \ref{critical curve},
we also prove that, no matter how large advection and diffusion rates are,
the specialist predators can invade successfully as long as they maintain a small mortality rate.

We also investigate the influence of advection and diffusion on the asymptotic profiles of the positive steady state solutions of systems \eqref{predator prey patch} and find that
diffusion and advection can dramatically affect the spatial distribution of species.
In Theorems \ref{UV q infty},
we demonstrate that two species will coexist and concentrate at the patch $n$ for a large positive advection.
This implies that the advection does not eliminate coexistence of species but significantly affect the distribution of species.
The intuitive biological explanation is that predators always keep pace with the prey, allowing them to successfully invade and coexist with the prey at the patch $n$ as flow speed increases.
In Theorems \ref{UV d1 infty} and \ref{UV d2 infty},
we show that, if the diffusion rate of the prey (resp. the predator) is large,
the species will coexist.
%in closed advective environments with a homogeneous distribution for the prey (resp. the predator).

If the prey has a small diffusion rate,
in Theorems \ref{critical curve} and \ref{properties of critical curve},
we prove that the specialist predators can always invade successfully as long as the mortality rate is suitably small.

If the specialist predators have a small diffusion rate,
they can successfully invade but with a small population density which tends to zero as the diffusion rate goes to zero. The biological explanation is that the specialist predators cannot catch up with the prey if they do not move fast (see Theorem \ref{properties of critical curve}(5) and \ref{UV d2 0}).

We end this section by proposing several interesting problems that deserve further consideration.
The first one is that
whether there is the uniqueness of positive steady state solutions of system \eqref{predator prey patch}.
Note that the arguments of deriving the uniqueness
as in Step 3 of \cite[Theorem 3.1]{NWW}
or the proof of \cite[Lemma 3.3, Theorem 3.4]{NHW}
do not work in our model \eqref{steady state} here.
For our model, a probable method to obtain the uniqueness is to
verify the invertible of the coefficient matrix.
However, due to the complexity of the coefficient matrix,
it is difficult to calculate it's determinant.
This requests us to further explore some new methods
to calculate the determinant or to consider the uniqueness.
Another problem concerns the dynamics of the specialist predator-prey model \eqref{predator prey patch} in open advective patchy environments.
Also, how does the dynamics change for the generalist predator-prey
patchy system ?
We leave these challenging problems for future investigation.

\end{document}